\documentclass[article]{interact}

\usepackage{epstopdf}
 
\usepackage[longnamesfirst,sort]{natbib}
\bibpunct[, ]{(}{)}{;}{a}{,}{,}
\renewcommand\bibfont{\fontsize{10}{12}\selectfont} 

\usepackage{amssymb}
\usepackage{amssymb}
\usepackage{graphicx}
\usepackage{url}
\usepackage{booktabs}
\usepackage{comment}
\usepackage{bbold}
\usepackage{placeins}
\usepackage{pdfpages}
\usepackage{adjustbox}
\usepackage{enumerate}
\usepackage{mathtools}
\usepackage{longtable}
\usepackage{blkarray}
\usepackage{caption}

\theoremstyle{plain} 
\newtheorem{theorem}{Theorem}[section]

\newtheorem{proposition}[theorem]{Proposition}
\theoremstyle{definition}

\newtheorem{example}[theorem]{Example}
\theoremstyle{remark}

\begin{document}

 \title{Using infinite server queues with partial information for occupancy prediction}

\author{
\name{Nikki Sonenberg\textsuperscript{a}, 
Victoria Volodina\textsuperscript{b},
Peter~G.~Challenor\textsuperscript{b,c},
Jim~Q.~Smith\textsuperscript{c,d},
\thanks{CONTACT N. Sonenberg. Author. Email: nikki.sonenberg@bristol.ac.uk}}
\affil{
\textsuperscript{a}University of Bristol, United Kingdom;
\textsuperscript{b}University of Exeter,  United Kingdom; 
\textsuperscript{c}The Alan Turing Institute, United Kingdom; 
\textsuperscript{d}University of Warwick, United Kingdom.}
}

\maketitle

\begin{abstract}
Motivated by demand prediction for the custodial prison population in England and Wales, this paper describes an approach to the study of service systems using infinite server queues, where the system has non-empty initial state and the elapsed time of individuals initially present is not known.
By separating the population into initial content and new arrivals, we can apply several techniques either separately or jointly to those sub-populations, to enable both short-term queue length predictions and longer-term considerations such as managing congestion and analysing the impact of potential interventions. The focus in the paper is the transient behaviour of the $M_t/G/\infty$ queue with a non-homogeneous Poisson arrival process and our analysis considers various possible simplifications, including approximation.  We illustrate the approach in that domain using publicly available data in a Bayesian framework to perform model inference. 
 \end{abstract}


\section{Introduction}

This work is motivated by the problem of predicting short and longer-term implications of policy changes on the custodial elements of the prison system in England and Wales.
The model described here was developed following consultation with the Ministry of Justice (MoJ) to add to their methods of forecasting the prison population, to help analyse the implications of changing external factors accounting for the prison population such as  government guidelines and sentencing policies, and to consider the uncertainty involved in the model and its predictions. 

The nature of the prison system is such that arrivals can’t be turned away, hence infinite server queueing models are directly applicable as they support the assumption of no queueing delay for service.
While such models have been widely used in modelling service systems, including call centres \citep{Ibrahim2016b}, hospital staffing \citep{Pender2016}, software reliability \citep{Huang2008} and insurance claims \citep{Cheung2019}, the assumptions relevant to our scenario lead us to consider some less well-known results from the queueing systems literature and discuss how they can be useful in this setting.

In this domain, matters of capacity management and overcrowding at individual institutions have to be handled by adjustments involving medium term system-wide considerations such as sentencing patterns, parole guidelines and the use of community service arrangements. This intent to support policy makers leads to a focus on the \emph{strategic} level of decision support~\citep{grieco2021,hulshof2012taxonomic}, and to some considerations on the development of  models expressed in terms that are interpretable by policy makers and can enable `what if?' studies, including quantification of the impact of prospective policy change~\citep{Bravo2019,Dong2020,kegel2017whatif,petris2009,tuominen2022forecasting}. 
 
The value of decision support tools that can analyse the impact of interventions linked to policy change has been demonstrated in care pathways in health services \citep{demir2017enabling}. While modelling flows through a prison system has not been widely studied, there has been some work using queueing models to study the relationship between policy changes and prison occupancy. For example,  \citeauthor{Usta2015} (\citeyear{Usta2015}) used 
a queueing network model of the jail system and a simulation approach to study the optimal mix of pretrial release and forms of sentencing to minimise
the amount of recidivism, subject to constraints on the available prison occupancy; and
\citeauthor{Master2019} (\citeyear{Master2019})  used a $M/M/c/c$ queueing system to assess performance of alternative pretrial release policies, and of sentences with a split of custodial and supervised outcomes.

Using public data, we study a single phase of a prisoner's journey through the prison system, as it is well out of scope to model the full system.  Our analysis is both informed by, and limited by, the availability of service system data:  the size of the prison population is collected and recorded on a monthly basis, but the time served and the remaining length of stay (service times) of those individuals are not. Indeed for custodial sentence admissions, even within offence types sentence lengths differ, and there is a difference between a court imposed sentence length and the actual service time, so the model cannot assume remaining service times for those individuals present at a given time, even if their formal sentence lengths are known. This contrasts with, for example, work on bed demand in an intensive care unit that considered existing patients as well as arrivals, but could use knowledge of the length of stay of existing patients \citep{Pagel2017}. 

 Queueing model analyses typically assume the system begins in an empty state \citep{Li2019} and then, assuming a Poisson arrival process, the well-known results of \cite{Eick1993} mean that the queue length exhibits a Poisson distribution with a mean derivable from properties of the service time distribution. 
The scenario motivating the work in this paper involves the less well-studied situation where
there are $n > 0$ individuals initially present and where the elapsed time at $t = 0$ of each individual is not known~\citep{Weber2005,Goldberg2008,Aras2017}.
In this case, the departure process is no longer Poisson, but the queue dynamics can be analysed as a combination involving those already in service at $t = 0$ and those who subsequently arrive, and the departure process is the superposition of a binomial and a Poisson process \citep{Weber2005}.
This separation into initial content and  new input allow these sub-populations to be analysed jointly or separately. 

As pointed out by \cite{Pagel2017}, in contrast to much research involving applications of queueing systems that rely on steady state distributions, when considering  the use of models for informing short term consequences associated with operational decision making, results involving transient distributions of queueing systems become relevant.

As highlighted in a recent review of major healthcare applications of infinite server queuing models \citep{Worthington2020}, time-inhomogeneous infinite server models have been used both for predictive modelling (e.g., ward capacity planning \citep{Bekker2010}) and for investigating the impact of policy changes (e.g., the introduction of specialised treatment centres for specified illnesses \citep{Utley2008}). Our ambition to support the analysis of both short and longer-term policy change means we also consider time-varying arrivals.

Another key choice in formulating a queueing theory model is the assumptions regarding service time. The available data suggest a heavy tailed service time \citep{offenderstats}. Hence, we are led to the $M_t/G/\infty$  model as it is well recognised that there is an effect on the performance of the queue of the tail of the distribution as it becomes less exponential \citep{Goldberg2008}.
 
To employ our model as a prediction tool, we must also consider the uncertainty involved as we are considering behaviour beyond the sample of the original data set used to estimate the model parameters \citep{Gans2003}.  Incorporating parameter uncertainty as part of model inference is an important step to avoid overconfidence in our results \citep{Aktekin2011}. We consider parameter uncertainty involved in the prediction of the system size using estimates from aggregate historical data. Techniques typically rely on summary observable data such as queue lengths, visit counts and response times, but perform inference in different ways \citep{Spinner2015}. A common approach is by \cite{Jongbloed2001} using a Poisson mixture model and a recent survey on forecasting of the arrival process is by \cite{Ibrahim2016b}. We employ a Bayesian framework to perform  model inference and prediction  \citep{Aktekin2011, Xie2014}. The Bayesian framework allows us to incorporate expert knowledge about the system into prior distributions of model parameters, which is particularly important since we have limited historical data available to us \citep{OHagan1998}.

In summary, we study the transient behaviour of the time-varying infinite server queue, $M_t/G/\infty$, fed by a non-homogeneous Poisson arrival process whose occupancy is observed at discrete points in time, but the time in service to that point is not known.
The contributions of this paper are: (i) the novel synthesis of results from several authors about transient and stationary behaviour of the $M_t/G/\infty$ queue; and (ii) application of the approach, using Bayesian inference, to the real-world domain of prison occupancy -- a domain that has not been well-studied in the literature.

The structure of this paper is as follows: in Section 
\ref{sec:prison} we describe the motivating application; Section \ref{observedqueuesection} outlines relevant results from the literature for analysing an observed $M_t/G/\infty$ queue; Section \ref{sec:param} presents a Bayesian framework for the estimation of the model parameters using domain data to illustrate the use of the model to both short-term and longer-term predictions, and includes a discussion of the mathematical assumptions of the model. In Section \ref{concludingremarks} we provide concluding remarks, with comments comparing the presented queueing theoretic approach with time-series based estimation methods. 
The Appendices contain further information about the prison system (Appendix \ref{app:prisons}), use of the theory (Appendix \ref{app:steadystate}), and some examples of how analysis with our model compares with using a time-series based ARIMA model (Appendix~\ref{app:comparison}).

 \section{Motivating application: Prison occupancy}
\label{sec:prison}

The study in this paper was produced following consultation with the Ministry of Justice (MoJ).
We briefly describe the custodial elements of the prison system, that is, those that require accommodation, with more details in Appendix \ref{app:prisons}. Data on the prison population is collected and managed by MoJ and the HM Prison and Probation Service (HMPPS).  Statistics are regularly released as well as projections of the prison population \citep{OffenderManage2019}.

Attributes of the application domain that guided our modelling choices include: a large system of multiclass arrivals with a high offered load, no abandonments,  input parameters that are subject to change and operation over a long time scale.  Further, the  domain data suggests the use of a non-exponential service time distribution and that a stationarity assumption is reasonable over short time frames, which allowed us to take advantage that several questions of interest have more tractable solutions under the assumptions of a stationary model.  

Factors such as the conviction rate (the proportion of those arriving to court that are convicted and sentenced) and the custody rate (the proportion of those sentenced that are given custodial sentences) influence the sentenced population.
Hence the size and composition of the prison population is subject to policy and legislation changes, for example, changes in Home Office (government department) resources that can affect charge rates and modifications to the sentencing guidelines (MoJ, \citeyear{PrisonProjections2019}). 
Patterns in the published data
illustrate how policy and legislative changes have had subsequent impacts on prison occupancy.
Hence, from a policy maker's perspective, being able to adjust model parameters to allow, for example, for a more serious mix of offence groups coming before the courts reflects the importance of reviewing model parameters over time.  

\begin{figure}[h!]
\begin{center}
\includegraphics[  width=.75\textwidth]{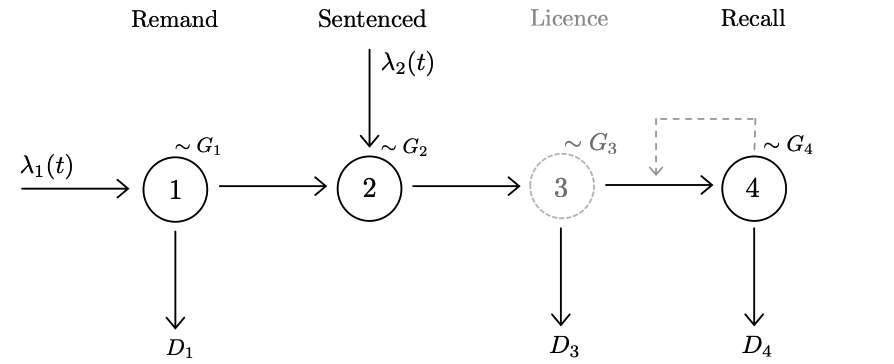}
\end{center}
\caption{Phases of the prison system}  
\label{model1sketch}
\end{figure}

Of course, describing the dynamics of the prison system is beyond the scope of this paper, but from a modelling perspective it requires only some reasonable assumptions to treat the flow of prisoners as an $M_t/G/\infty$ queue \citep{Schwarz2016}. Figure \ref{model1sketch} displays our model of a simplified prisoner journey  with the prison population divided into three main holding phases (displayed percentages are as of June 2019) (MoJ, \citeyear{PrisonProjections2019}): (i) on {remand} (11\%), (ii) sentenced prisoners (79\%) and (iii) on {recall} (9\%). Prisoners within the licence phase are in the community. 
A broader view of the prison system and its constraints would include, for example, the number of offenders on probation, staffing resources required for supervision, demands on the courts and parole hearing frequencies \citep{Crowhurst2016,MoJStory2016,OffenderManage2019}.

\section{An observed $M_t/G/\infty$ queue}
\label{observedqueuesection}

Motivated by obtaining a prediction of the population given  partial information, 
 we describe results applicable to an infinite server system with Poisson arrivals and general service times observed at time $\tau > 0$, under the assumption that we have no information on when the $n$ individuals present at this time each began their service, namely an \emph{observed $M_t/G/\infty$ queue}.  
 We define the infinite server queue in  Section \ref{mtginfsub} and then present results for the conditional distribution for the observed queue in Section \ref{subsec:conditional}. 
 As the $n$ individuals initially observed at time $\tau$ complete service, this sub-population will go to zero as $t \to \infty$. Arrivals after time $\tau$ will occur according to the Poisson dynamics described in Section \ref{mtginfsub}.  These results are used in our empirical study in Section \ref{sec:param}.

\subsection{The $M_t/G/\infty$ queue}\label{mtginfsub}
The $M_t/G/\infty$ queue is a service system in which individuals arrive according to a non-homogeneous Poisson process with rate function $\lambda(t)$, for $-\infty<t <\infty$ and where the service times are independent and identically distributed  (i.i.d.) and independent of the arrival process \citep{Eick1993}. Let $S$ be the service time and denote by $G$ its cumulative distribution function (cdf) and $g$ its density.
Assume $E[S]<\infty$, $\lambda(t)$ is bounded and integrable and define the associated random variable $S_e$, the stationary excess of the service time,
 with cdf $G_e(t) = P(S_e \le t)$ for $t \ge 0$,   
\begin{align} \label{defnexcessdistn}
G_e(t) &= \frac{1}{E[S]} \int_0^t G^c(u) du,
\end{align}
where $ G^c(u) = 1- G(u)$.

Let $Q(t)$ represent the number of busy servers at time $t$ and let $m(t) = E[Q(t)]$.
\begin{theorem}\label{masseysingletheorem}   \citep[Theorem 1]{Eick1993} 
For an $M_t/G/\infty$ queue that was initially empty at $t = -\infty$, for each $t$, $Q(t)$ has a Poisson distribution with mean  
\begin{align}
m(t) &= E \left [ \int_{t-S}^t \lambda(u) du \right ]
= E \left [ \lambda(t-S_e)\right ] E[S]. \label{massey1993theorem11}
\end{align}
The departure process is a Poisson process with time dependent rate function $\lambda^{-}(t)$, where
\begin{align}\label{departuretheo}
\lambda^{-}(t) & = E \left [ \lambda(t-S)\right ].
\end{align}
For each $t$, $Q(t)$ is independent of the departure process in the interval $(-\infty, t]$. 
\end{theorem}
 
The impact of the service time beyond its mean can be seen from $E[S_e] = \frac{1}{2}{E[S] (c_s^2 +1)}$ where $c_s^2 =Var(S)/E[S]^2 $  is the squared coefficient of variation (SCV).

 For $\lambda(t) = \lambda$, the approach to the steady state is given by 
\begin{align}
{m}(t) &= \lambda G_e(t) E[S],\label{stepfunctionmean}
\end{align}
where the steady state has the insensitivity property, as $m(\infty) = \lambda  E[S]$. Similarly, the transient behaviour of a stationary model that has been terminated, that is, for $t<0$, $\lambda(t) = \lambda$ and zero otherwise, is
$ {m}(t) = \lambda G^c_e( t) E[S]. $  

In an empty system, the approach to steady state is seen from Equation \eqref{stepfunctionmean}. For a non-empty system observed at time $\tau$, we note a result by \citet{Mandjes2011}  who analysed the approach to steady state 
identifying a function that behaves as the difference between the transient and stationary distributions in the limit, and of relevance in the following Section \ref{subsec:conditional}.

\begin{example}\label{ex:pareto}
We use the Pareto distribution as it exhibits the heavy tailed non-exponential survival times observed in the prison domain, and we draw upon this later. For $G \sim Pa(\theta,\alpha)$ with  shift parameter $\theta>0$ and shape parameter $\alpha>0$, for $x \ge 0$,  
$ G^c(x)= \theta^{\alpha} (x+\theta)^{-\alpha}$.
The high variability of $G$ is indicated by the fact that the tail decays as a power instead of exponentially.  
For $\alpha >1$, $E[S] = \theta(\alpha-1)^{-1}$, otherwise if $\alpha \le1$ then $S$ has infinite mean and $G^c(x)$ is not a directly integrable function.
For $\alpha >2$, $ Var[S] =  \theta^2 \alpha(\alpha-1)^{-2} (\alpha-2)^{-1}$, for $1<\alpha \le2$  the variance is infinite, and for $\alpha$ otherwise the variance is undefined. For $\alpha>1$,
$G^c_e(x) =     \theta^{\alpha-1}(x+\theta)^{1-\alpha}$.
 If $\alpha >2$ then the SCV is
$c_s^2 =  \alpha(\alpha-2)^{-1}$, then $E[S_e] =  \theta(\alpha-2)^{-1}$.
\end{example}

\subsection{Conditional distribution}\label{subsec:conditional}
Denote by $\tau$ the observation time, by $\tau+\delta$ the prediction time, and define the  vector of the past arrival intensity,
 $\hat{\lambda} = \{\lambda(t) : 0 \le t \le \tau\}$.  For a stochastic process $f(t)$ cut at an observation time $\tau$, define the past process
$ \hat{f}(t) =   f(t)$ if $t <\tau$ and $\hat{f}(t) = 0$ if $t \ge \tau$, and define the future process as $ \check{f}(t) =0$, if $t <\tau$ and $ \check{f}(t) =f(t)$ if $t \ge \tau$. 
The past process can be thought of as an $M_t/G/\infty$ process in which arrivals are terminated at time $\tau$, so we are able to draw on results by \citet{Goldberg2008} who, motivated by a model of a two-stage item inspection process, studied the distribution of the last departure time from a queue of a terminating arrival process.
 
\begin{figure}[h!]
\begin{center}
\includegraphics[width=.5\textwidth]{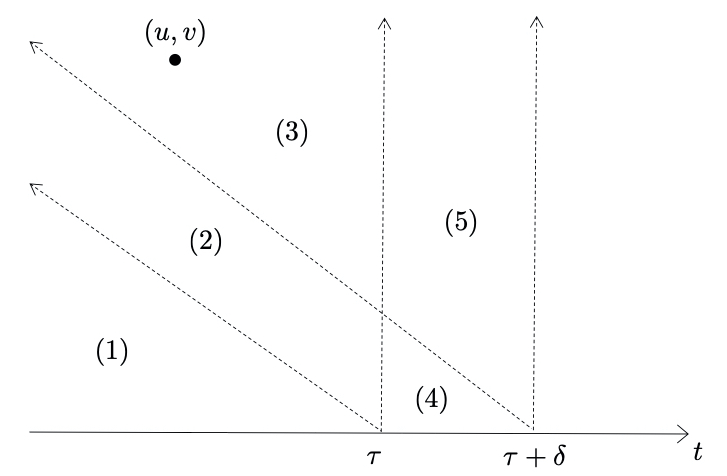}
\end{center}
\vspace{-1em}
\caption{Regions describing arrival and service pairs $(u,v)$.} 
\label{webergraphproofdiag}
\end{figure}

 For the analysis of the observed queue, we separate the contributions of the sub-populations of those observed at time $\tau$ and those arriving later. It is useful to consider the regions describing arrival and service pairs as depicted in Figure \ref{webergraphproofdiag}, where for each individual arriving to the system, a point is placed at $(u,v)$, with the $u$-axis recording arrival times and $v$-axis recording service times. For example, Region (1) corresponds  to individuals arriving and departing by $\tau$, and the region $A_{(2 \cup  3)} = \{(u,v)  \mid \ u \le \tau, u+v \ge \tau \}$ corresponds to individuals arriving on or before time $\tau$. 
Denote by $N(A)$ the number of arrival-service pairs in a region $A$. As per Theorem \ref{masseysingletheorem}, the Poisson arrivals and general service times generate a Poisson process on the plane with the intensity of a point occurring at $(u,v)$ being $(\lambda(u), g(v))$.  As independent splitting of Poisson processes results in Poisson processes, the numbers of points in two disjoint regions are independent Poisson random variables. For example, consider $A_{(3)} = \{(u,v)  \mid \ u \le \tau, u+v \ge \tau+\delta \}$, corresponding to individuals arriving by $\tau$ and present at $\tau+\delta$. Then the proportion of individuals present at time $\tau$ whose remaining service time from that point is at least $\delta$, is clearly $N(A_{(3)})/N(A_{(2 \cup 3)})$. This geometric depiction is captured in the following result.

 \begin{theorem} \citep[Theorem 2.1]{Goldberg2008} \label{remainingtheogoldberg}
 Conditional on there being $n$ individuals in the system at time $\tau$, the remaining service times are i.i.d., each distributed as a random variable $X_\tau$  
 with ccdf
 \begin{align}
G^c_\tau(x) &= P \left ( X_\tau > x \right ) \label{ccdfremain}= \frac{1}{\nu_\tau} \int_{-\infty}^\tau \lambda(u) G^c(\tau+x-u) du.
\end{align}
where mean $\nu_\tau$ is given by
 \begin{align}
\nu_\tau =  \int_{-\infty}^\tau\lambda(u) G^c(\tau-u) du = 
 \int_{0}^\infty \lambda(\tau - u) G^c(u) du. \label{meannumbertau}
\end{align} 
 \end{theorem}

  It directly follows that if $\lambda(t) = \lambda$ for $t \ge 0$, and  $\lambda(t) = 0$ for $t < 0$, then $\nu_\tau = \lambda E[S]G_e(\tau)$ and 
\begin{align}
G_\tau^c(x) = \frac{G_e(\tau+x) - G_e(x)}{G_e(\tau)}.
\end{align}
 
The above result provides an expression for the remaining service time distribution in Equation \eqref{ccdfremain}, which is required to calculate the conditional distribution in the following result. 

\begin{theorem}\citep[Theorem 6]{Weber2005} 
The random variable $Q(\tau+\delta \mid  \tau)$ with $Q(\tau) = n$ can be expressed as
\begin{align} \label{rvQtaudelta}
Q(\tau+\delta|\tau) =  Bi\left(n, p_{\tau}(\delta)  \right ) + Po\left (\check{m}(\tau+ \delta) \right) ,
\end{align}
where
\begin{align}
\check{m}(\tau+\delta) = & E[\check{\lambda}(\tau+\delta - S_e)]E[S] =\int^\delta_0 \lambda(\tau+ \delta - u)G^c(u) du, \label{poisonweberintegral}
\end{align}
 where $p_{\tau}(\delta) = G^c_\tau(\delta) $ is given in Equation (\ref{ccdfremain}), and $Po(\cdot)$ and $Bi(\cdot, \cdot)$ are the Poisson  and Binomial random variables, respectively. 
 \end{theorem}

  The process $Q(\tau+\delta)$ can be written as
$Q(\tau+\delta) = \hat{Q}(\tau+\delta) +  \check{Q}(\tau+\delta)$, where  $\hat{Q}(t)$ has arrival rate $\hat{\lambda}(t)$ and  $\check{Q}(t)$ has arrival rate $\check{\lambda}(t)$.   
As constructed above, $Q(\tau+\delta\mid\tau)$ will be the number of points in regions $(3 \cup 5)$ of Figure \ref{webergraphproofdiag}. $N(A_{(5)})$ has Poisson distribution with mean given in Theorem \ref{masseysingletheorem} and the result for $\check{Q}(\tau+\delta | \tau)$ is given by Equation~(\ref{poisonweberintegral}). The distribution of $N(A_{(3)})$ is binomial with $n$ trials and parameter $p_\tau(\delta)$ given in Equation \eqref{ccdfremain}.

Depending on the size of $\delta$ relative to $\tau$, either existing or new arrivals can dominate the prediction. The calculations can reveal the contribution of each component to future demand requirements. For some quantities of interest the two sub-processes $Q(t) = (\hat{Q}(t), \check{Q}(t))$, from the expression in Equation \eqref{rvQtaudelta} can be studied separately. The results extend easily to multiple classes, $i$, $\{Q_i(\tau) = n_i\}$  as each class is treated independently. 
\begin{proposition}
The conditional distribution is
\begin{align}\label{transient_dist_thm}
&P\left [Q(\tau+\delta) = y\mid Q(\tau) = n \right]  
=  \sum_{k=0}^{\min{\{n,y\}}}P \left[Bi\left(n, p_{\tau}(\delta)  \right ) = k] P[Po\left (\check{m}(\tau+\delta) \right) = y-k \right ],
  \end{align} with mean
$    E\left [Q(\tau+\delta) \mid Q(\tau) = n \right] = n p_{\tau}(\delta) + \check{m}(\tau +\delta)$,
    and variance
$    Var\left[Q(\tau+\delta) \mid Q(\tau) = n) \right] = n p_{\tau}(\delta)(1-p_{\tau}(\delta) )+ \check{m}(\tau +\delta)$.
\end{proposition}
 
The departure process of the observed $M_t/G/\infty$ queue at time $\tau+\delta$ will be the superposition of a Poisson process with intensity 
$\lambda^-(\tau +\delta) = \int_{0}^\delta \lambda(\tau+ \delta - u) g(u) du$
and a binomial process with parameters  $n$ and  $1- p_{\tau}(\delta)$ \citep{Weber2005}. A consequence of which is that the Poisson-in-Poisson-out feature of the unobserved $M_t/G/\infty$ system does not hold.

Under a high load, the binomial process will be the superposition of several point processes and will be approximately Poisson: for large $n$, 
    \begin{align} \label{eq:approxhighloadpo}
P[\tilde{Q}(\tau+\delta)=y\vert\tilde{Q}(\tau)=n] 
&=P[Po(np_{\tau}(\delta)+\check{m}(\tau+\delta))= y].
\end{align}  
Thus considering the system began with an initial population drawn from a Poisson distribution with mean $n$ and we employ this approximation in Section \ref{sec:param}.

 \subsubsection{Last departure time}\label{lastdeparture}

The behaviour of the decay of the initial population $\hat{Q}(t)$ can be described by 
results of  \cite{Goldberg2008}.   Suppose $n$ individuals are observed at time $\tau$ and consider the time of the last departure. Define $M_n = \max\{X^1_\tau, \dots, X^n_\tau\}$, where each $X^i_\tau$ is distributed according to Equation \eqref{ccdfremain}, then 
$ P \left [ M_n \le x\right] = G_{\tau}(x)^n$.
Now consider the case where the arrival process is terminated at time $\gamma$, but where an observation is not made. Following \citet{Goldberg2008},  let $D$ be the last departure time and define $T = (D-\gamma)^+$, the remaining time after $\gamma$ until the last departure. To determine the distribution of $T$, let $N = N_\gamma$ be a random variable with a Poisson distribution having mean $\nu_\gamma$, then $T \stackrel{d}{=} M_N.$
 
 For any random variable $Y$ with continuous cdf, let its quantile function be $q_Y \equiv q_Y(x)$ such that the number $P[T \le q_Y(x)]=x$. Write $f(x) \sim g(x)$ as $x \to \infty$ when $f(x)/g(x) \to 1$ as $x \to \infty$.

 \begin{theorem}\citep[Theorem 2.2]{Goldberg2008}\label{goldbergtheorem22} 
 For any $x >0$,
 \begin{align}
 P\left [ T \le x \right] = e^{-\nu_tG_\gamma^c(x)}  \label{goldberg1}.
\end{align}
Then as $x \to \infty$, $
P[T>x] \sim \nu_\gamma G_\gamma^c(x) \label{goldberg2},$
and for $e^{-\nu_\gamma} <x <1$,
\begin{align}
q_T(x) = q_{X_\gamma} \left ( 1 - \frac{\log(1/x)}{\nu_\gamma}\right ). \label{goldberg3}
\end{align}
The moments of which can be calculated by $E[T^k] = \int_0^\infty k x^{k-1}E[T>x] dx$.
 
\end{theorem}

\begin{figure}[h!]
   \centering
 \includegraphics[width=0.5\textwidth]{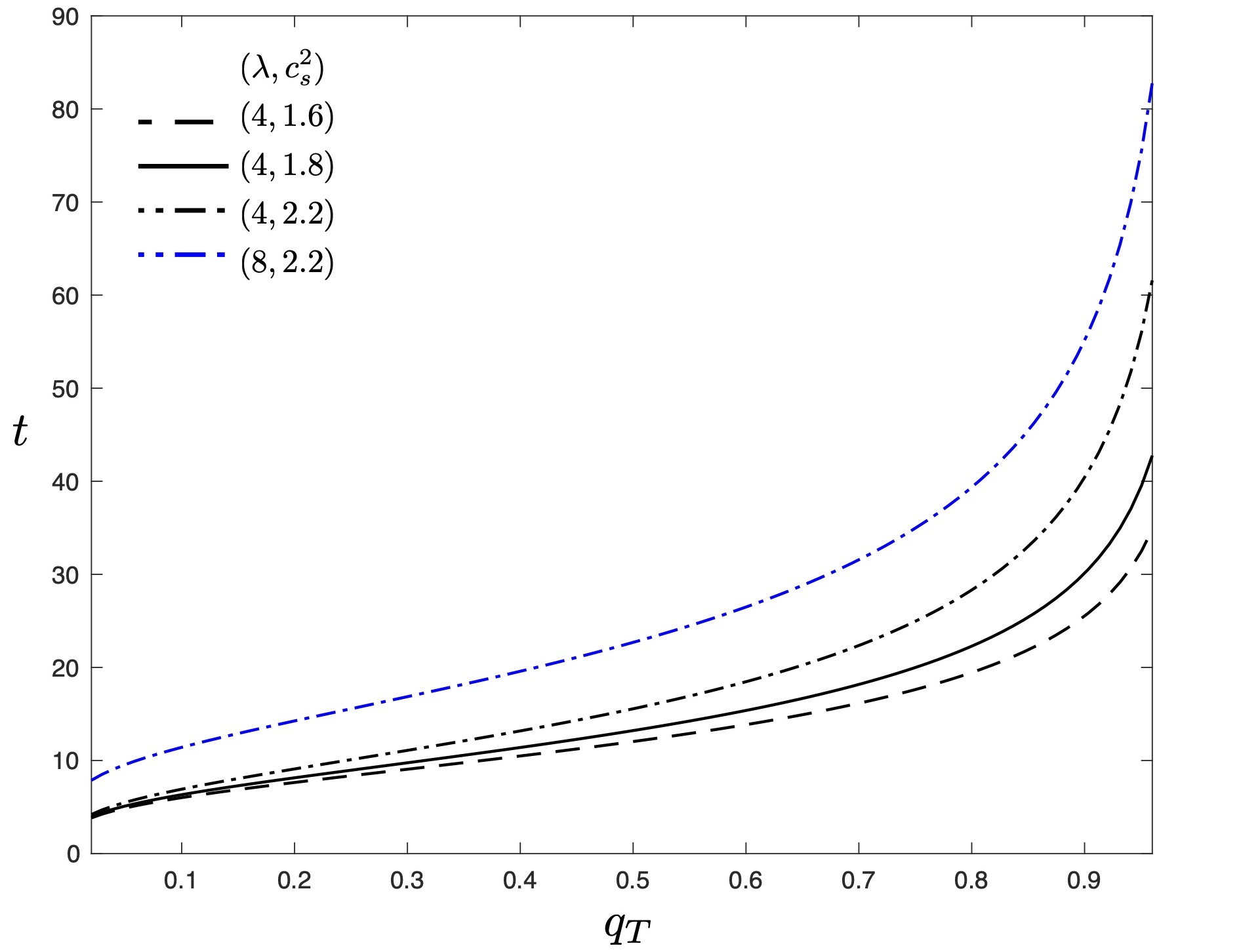}
  \caption{Example \ref{ex:pareto_quartile}. For fixed $E[S] = 3$, varying $(\lambda, c_s^2)$.  }\label{fig:example_quartile}
\end{figure}

\begin{example}\label{ex:pareto_quartile}
For $G \sim Pa(\theta,\alpha$), and with $\lambda(t) = \lambda$ for $ t \in (-\infty, \tau]$,
\begin{align}
q_T(x) = q_{X_\tau} \left ( 1 +\frac{(\alpha-1)\log(x)}{\lambda \theta}\right ),
\end{align}
with
\begin{align}\label{ex:pareto_quartT}
q_{X_\tau}(x)&=  \left ( \frac{\theta^{\alpha-1}}{ 1 - x   } \right)^{\frac{1}{\alpha-1}} - \theta.
\end{align}
In Figure \ref{fig:example_quartile} for  fixed $E[S] = 3$, we illustrate varying values $(\lambda, c_s^2)$. The effect of increasing the SCV is seen in the quantiles of the last departure time $T$. 
These results can be employed in the application domain to study the decreasing  content of a prison population with no new arrivals.
 
\end{example}

 \section{Empirical study and simulations}\label{sec:param}
 
The statistical inference of infinite server queues has been studied given various types of incomplete data: (i) queue length data \citep{Picklands1997,Goldenshluger2019}; (ii) unmatched arrival and service times \citep{Blanghaps2013};  
(iii) busy period data \citep{Hall2004}; and (iv) arrival and departure counts \citep{Li2019}. 

In this section, we present a method to generate predictions that consider the parameter uncertainty using Bayesian inference, considering queue length data from March 2015 to March 2019 on the Adult/Male sentenced population(MoJ, \citeyear{offenderstats}, Table 1.2b: Sentenced to immediate custody).
We refer to the MoJ documentation for details on counting processes, but we interpret this as the sentenced population and only consider data from 2015 due to changes in the reporting processes.

 We aim to provide a demonstration of the estimation approach for the model parameters using simulated monthly counts generated from published quarterly data, as the data is not sufficient to perform Bayesian inference about the system and model parameters. To generate monthly counts we: (i) divide the quarterly counts to obtain a monthly mean count; (ii) fit a smoothing spline to produce monthly predictions; (iii) add noise equal to the residual standard error from a linear model fit.

\subsection{Bayesian analysis of the sentenced population}\label{subsec:Bayes}
 
To specify the model, we adopt a linear arrival rate of the form $\lambda(t)=\beta_0+\beta_1t$ and assume that the service time distribution for each offence group is Pareto, and can be adequately estimated from historical data (over a much longer time interval than the prediction lead time).  
  Using public data (MoJ, \citeyear{crim_stats}, July 2018: June 2019) for the mean sentence length for each offence group,   we specify that the  service time is 50\% of the sentence length for this demonstration.

For a large system, define the mean function $M(\tau+\delta) = E[\tilde{Q}(\tau+\delta)]$ using Equation \eqref{eq:approxhighloadpo},  and Pareto$(\theta, \alpha)$ distribution as in Example \ref{ex:pareto}, we have
\begin{align}
v_{\tau}&=\frac{(\beta_0+\beta_1\tau)\theta}{\alpha-1}-\frac{\beta_1\theta^2}{(\alpha-1)(\alpha-2)}, \\
p_{\tau}(\delta)&=\theta^{\alpha-1}(\delta+\theta)^{1-\alpha}\Bigg[1+\frac{\beta_1\delta}{(\beta_0+\beta_1\tau)(2-\alpha)+\beta_1\theta} \Bigg],\\ \check{m}(\tau+\delta)&=\frac{(\beta_0+\beta_1\tau)}{1-\alpha}[\theta^{\alpha}(\delta+\theta)^{1-\alpha}-\theta]\nonumber\\
&+\frac{\beta_1}{(1-\alpha)(2-\alpha)}[\theta^{\alpha}(\delta+\theta)^{2-\alpha}-\theta^2-\delta\theta(2-\alpha)].
\end{align}

We specify prior distributions for arrival rate function and service distribution time parameters $\{\beta_0, \beta_1, \alpha, \theta \}$, where arrival data is used to specify informative priors for arrival rate function. We define a Bayesian model for the monthly numbers of arrivals with non-informative priors. 
Denote by $Q_a(\tau+\delta)-Q_a(\tau)$ the number of individuals that arrive in $[\tau, \tau+\delta$) and is Poisson distributed with mean $    \int_{\tau}^{\tau+\delta}(\beta_0+\beta_1u) du$. We extract the posterior samples of $\beta_0$ and $\beta_1$. Since both posterior densities are bell-shaped, we define normal priors for $\beta_0$ and $\beta_1$ centred on the mean of posterior samples and standard deviation equal to the scaled posterior sample standard deviation for the model. We use a scaled posterior sample standard deviation in order to fully explore the parameter space during Markov chain Monte Carlo (MCMC). We specify a uniform prior for $\alpha\sim\text{Uniform}[2.5, 10]$, where the lower bound avoids infinite variance and the upper bound ensures good convergence and mixing of the Markov chains. In practice, we observe that increasing the upper bound of $\alpha$ leads to non-identifiability.
We use RStan to perform full Bayesian statistical inference by adopting the Hamiltonian Monte Carlo (HMC) and no-U-turn samplers (NUTS) \citep{Carpenter2017}.

\label{subsubsec: Bayesian prediction}
We demonstrate long-term ($k$-step ahead prediction) and short-term (one-step ahead prediction) forecasts. The long-term forecasts are useful in informing long-term planning decisions in relation to the prison system capacity, whereas short-term forecasts could provide the insight needed for the day-to-day prison system operation. To generate predictions about the system behaviour, we perform the following steps:  \begin{enumerate}
\item Sample $(\beta_0^{(k)}, \beta_1^{(k)}, \alpha^{(k)}, \theta^{(k)})$ from the posterior distribution $[\beta_0, \beta_1, \alpha, \theta\vert\boldsymbol{n}^s]$ using RStan, where $\boldsymbol{n}^s=(n_1, \dots, n_{\tau})$ with $n_i, i=1, \dots, \tau$ being the number of individuals present in the system at time $i$.
\item For each $k=1, \dots, N$ compute $M^{(k)}(\tau+\delta)$  and sample $\tilde{Q}^{(k)}(\tau+\delta)$ from $Po(M^{(k)}(\tau+\delta))$.
\item Compute sample mean  $\mu_{\tilde{Q}}(\tau+\delta)=\frac{1}{N}\sum_{k=1}^N\tilde{Q}^{(k)}(\tau+\delta)$, which is our prediction.  
\item Compute sample standard deviation, 
$\sigma_{\tilde{Q}}(\tau+\delta)=$$\sqrt{\frac{1}{N}\sum_{k=1}^N\big(\tilde{Q}^{(k)}(\tau+\delta)-\mu_{\tilde{Q}}(\tau+\delta)\big)^2},
$
which represents uncertainty about our prediction.
\end{enumerate}

\subsection{Predictions and simulation results}\label{subsec:sims}

\begin{figure}[h!]
\begin{center}
\includegraphics[width=0.65\textwidth,height=0.38\textheight]{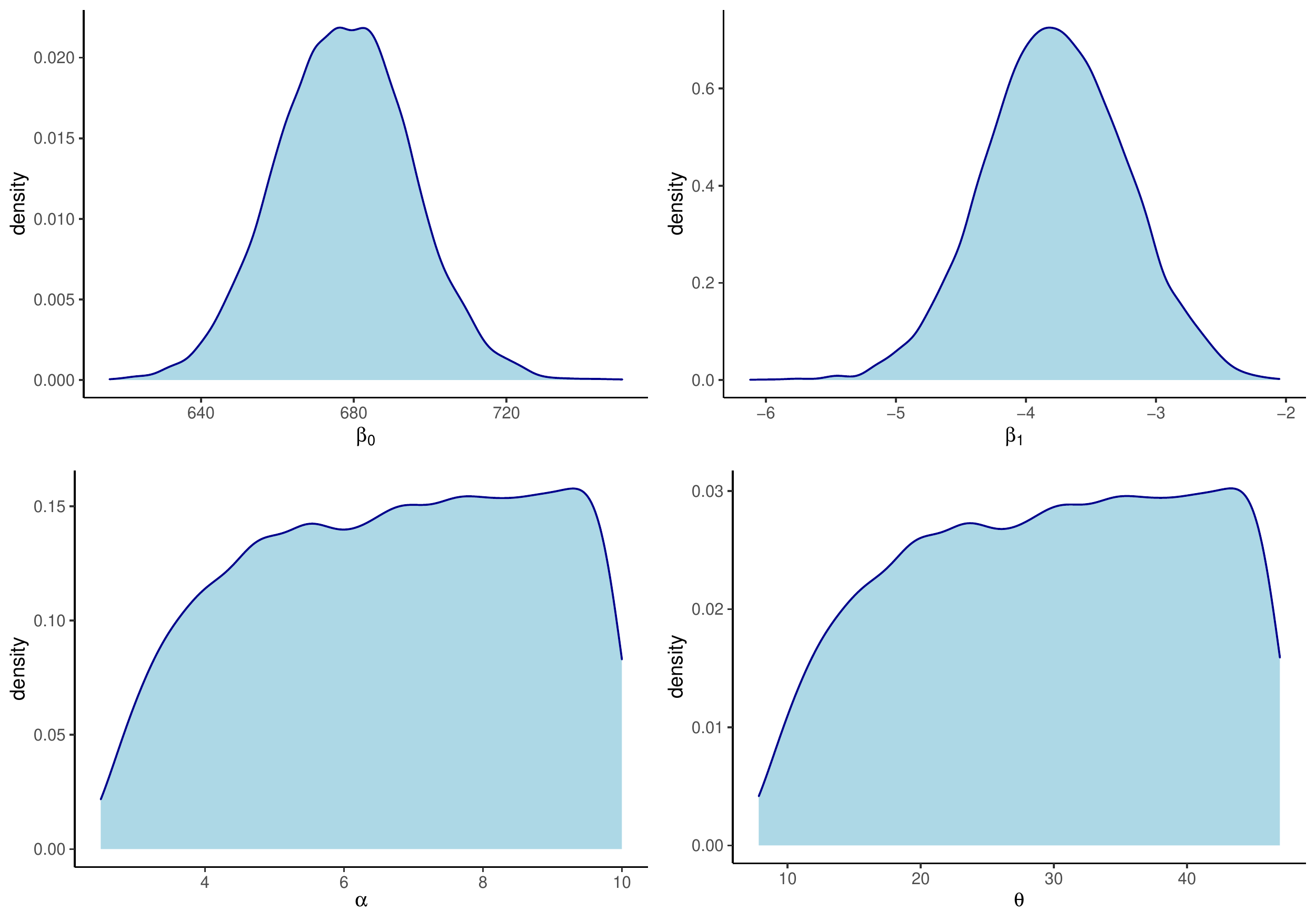}
\end{center}
\vspace{-1em}
\caption{Left to right: density plots of marginal posterior distributions for parameters $\beta_0$, $\beta_1$, $\alpha$ and $\theta$.}
\label{fig:PosteriorSamplesTheft}
\end{figure}

\begin{example} \label{ex:theft}
We demonstrate the Bayesian model specification to predict the number of prisoners in the Theft offence group where $E[S] = 5.22$ months. From the arrival count data, 
we obtain the posterior sample for $\beta_0$ with mean 1376.5 and standard deviation 9.7. Similarly, the mean and standard deviation of posterior sample for $\beta_1$ are -11.5 and 0.3 respectively. We then specify the priors: $    \beta_0\sim \text{Normal}(1376.5, 10\times 9.7)$,  $    \beta_1\sim \text{Normal}(-11.5, 10\times 0.3)$ and $ \alpha \sim \text{Uniform}[2.5, 10]$,
with $  \theta=5.22(\alpha-1) $. We set two Markov chains with 10,000 iterations for each chain (including warmup).

In Figure \ref{fig:PosteriorSamplesTheft}, we produce the density plots of marginal posterior distributions for model parameters. We observe that the posterior sample density of $\beta_0$ is bell shaped and centered around 677.4 and the standard deviation of posterior samples is 17.76. Similarly, the values of $\beta_1$ is centred around -3.77 with posterior standard deviation 0.54. We observe that under our model specification the number of arrivals in Theft Offence group gradually decreases over time. The distributions of posterior samples for $\theta$ and $\alpha$ are less informative, since we included the mean of service time in our prior specification.

In Figure \ref{fig:Prediction_Theft} we illustrate long and short term predictions from August 2019 to March 2020. We note that the short-term projections are more computationally expensive as in order to obtain a new observation, we update the posterior distribution by re-running a Stan programme. The left panel plot in Figure \ref{fig:Prediction_Theft} shows the long-term forecast (for 8 months). The number of offenders in Theft offence category is expected to decrease over time. Our predictions slightly underestimate the true values, however the true values lie within two standard deviation prediction interval. We observe that uncertainty about our projections increases with time. 
The right panel plot in Figure \ref{fig:Prediction_Theft} shows the short-term forecasts. The predictions are closer to the true values together with the slightly lower uncertainty about our predictions. 
To access the performance of the proposed model, we use the Root Mean Square Error (RMSE): $ RMSE=\sqrt{q^{-1}{\sum_{
\delta=1}^q (n_{\tau+\delta}-\mu_{\tilde{Q}}(\tau+\delta))^2} }$,
where a lower value indicates a better model performance.
The RMSE for long-term predictions is  53.08, whereas for short-term predictions is 20.9.
\end{example}

\begin{figure}[h!]
\begin{center}
\includegraphics[width=0.65\textwidth,height=0.23\textheight]{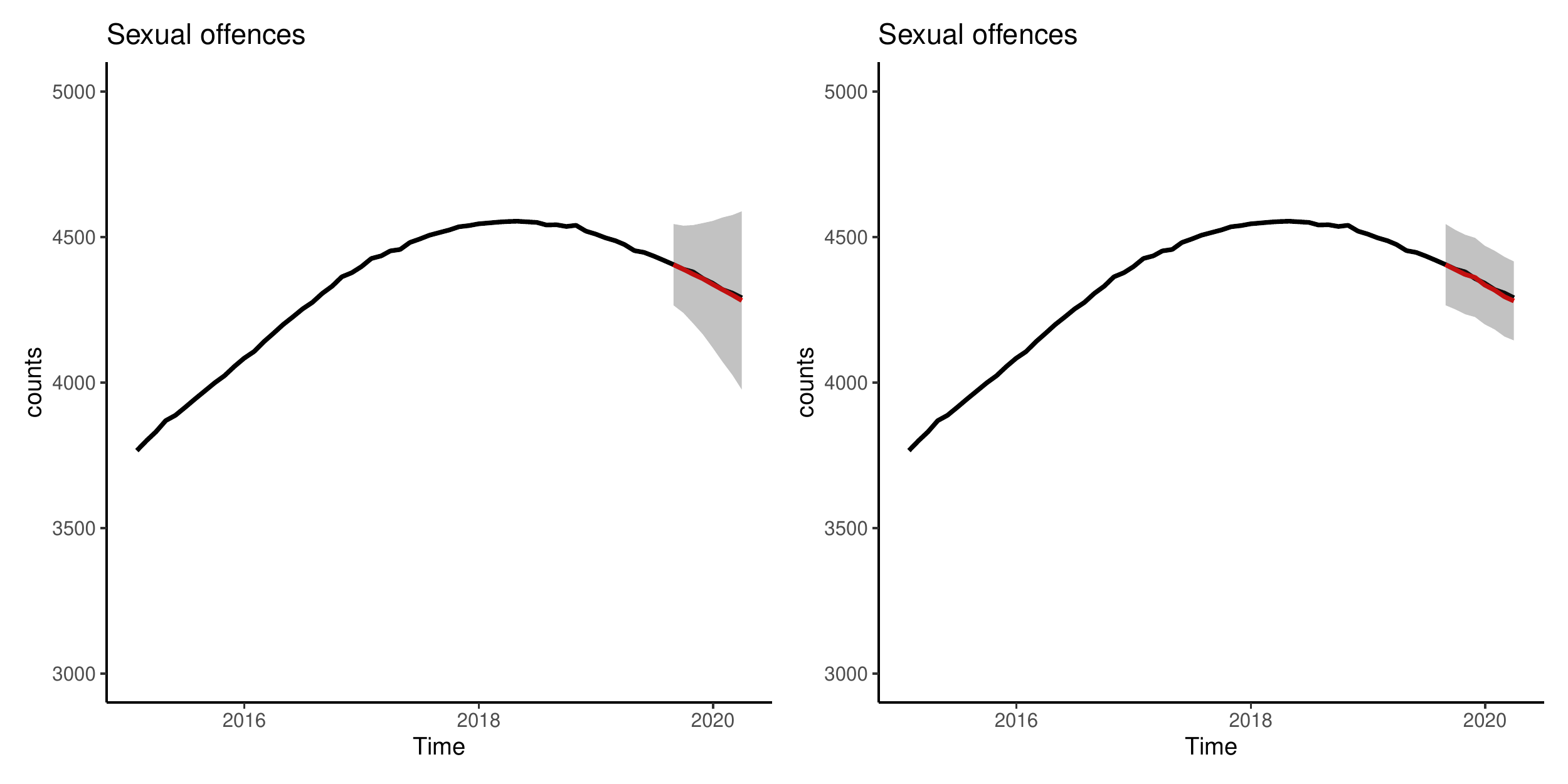}
\includegraphics[width=0.65\textwidth,height=0.23\textheight]{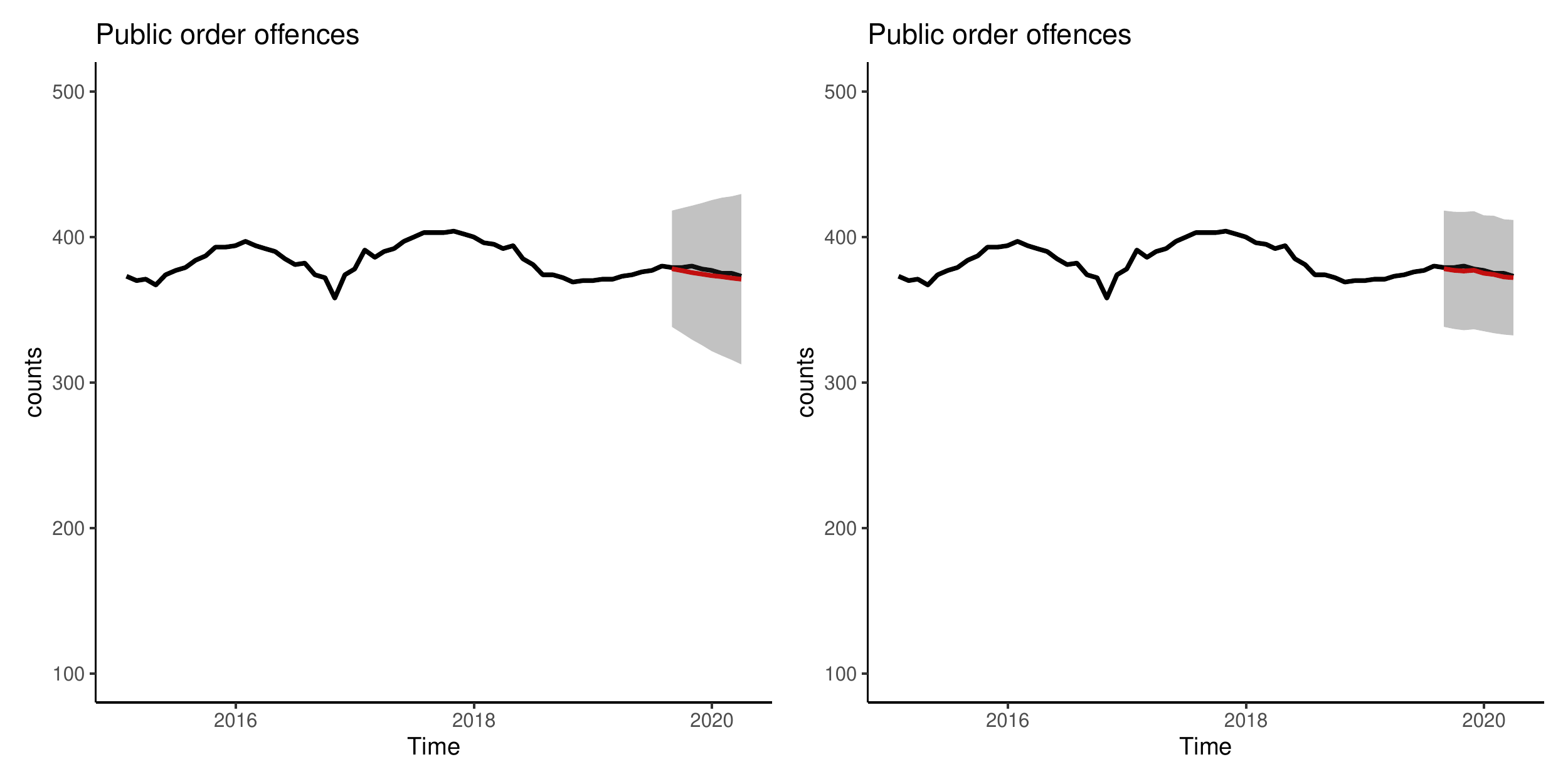}
\end{center}
\vspace{-1em}
\caption{Prediction of the number of prisoners in the Sexual offences (\textit{top row}) and Public order offences (\textit{bottom row}) groups from August 2019 to March 2020. \textit{Left panel:} long-term projections (red line) together with two standard deviation prediction interval (grey shaded region). \textit{Right panel:} short-term projections (red line) together with two standard deviation prediction interval (grey shaded region). The observed values are in black.}
\label{fig:Prediction_PublicOrder}
\end{figure}

\begin{example}\label{ex:public}
Adopting a  similar approach as in Example \ref{ex:theft}, we consider the number of offenders for two further offence groups: Sexual offences and Public order offences. The predictions for  Public order offences are illustrated in Figure \ref{fig:Prediction_PublicOrder}. 
For Sexual offences the RMSE for long-term forecast is 7.16, whereas for short-term forecast is 5.10. For Public order offences: the RMSE for long-term forecast is 2.97, whereas for short-term forecast is 1.89.
\end{example}

In the next example we reuse the posterior samples of model parameters derived in Examples \ref{ex:theft} and \ref{ex:public} to demonstrate how the model can provide insight into policy modifications.

\begin{figure}[h!]
\begin{center}
\includegraphics[width=0.6\textwidth,height=0.25\textheight]{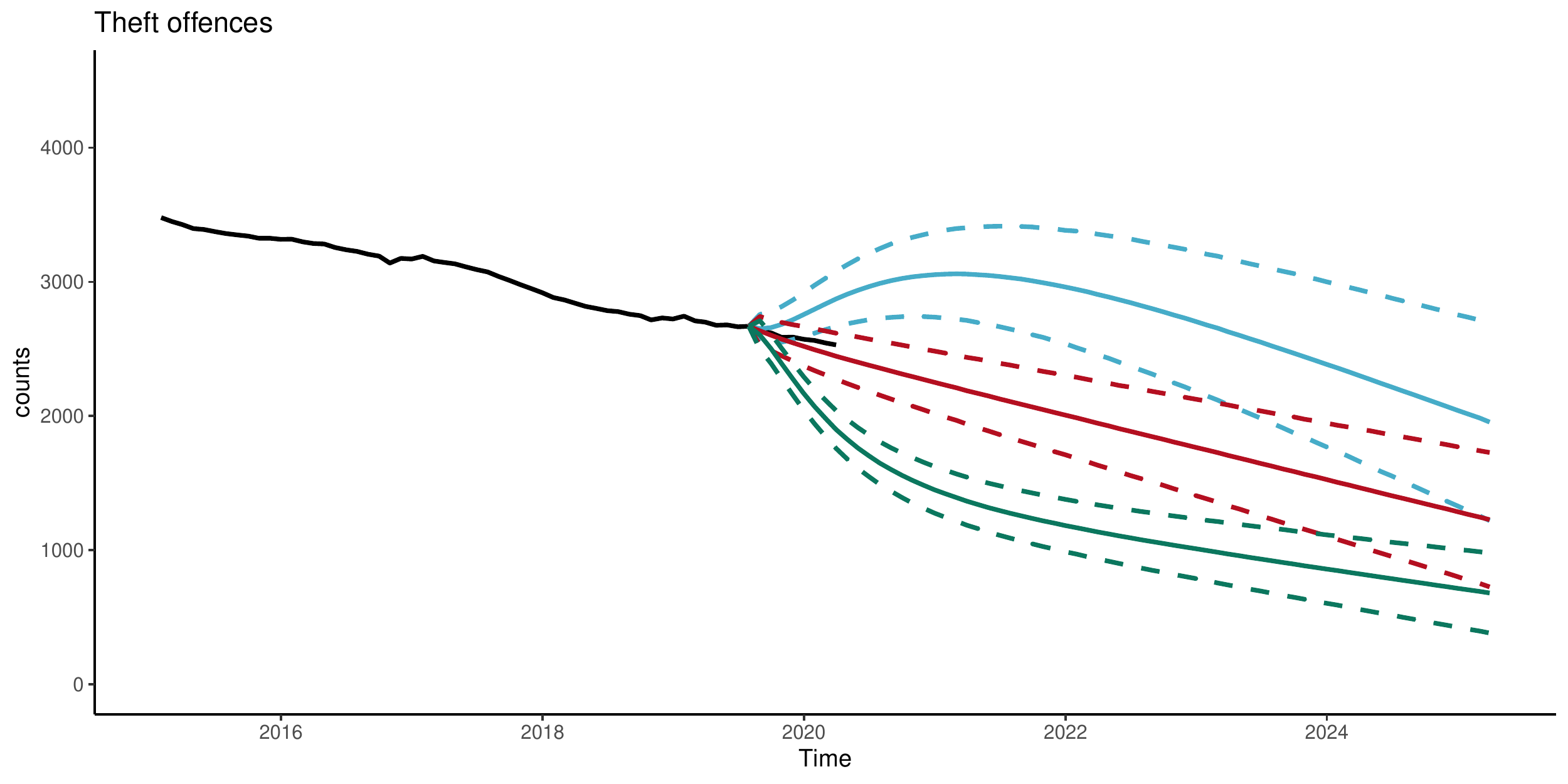}
\end{center}
\vspace{-1em}
\caption{Example \ref{ex:policymod}. Sentencing policy modification for Theft offences group where $E[S^o]=5.22$. The red line is associated with unchanged parameters, the blue line is associated with $E[S^\nu]=8$ and the green line with $E[S^\nu]=3$. Solid line is the predictor. Dashed lines represent two standard deviation prediction intervals.}
\label{fig:Theft_scenarios1}
\end{figure}

\begin{figure}[h!]
\begin{center}
\includegraphics[width=0.5\textwidth,height=0.25\textheight]{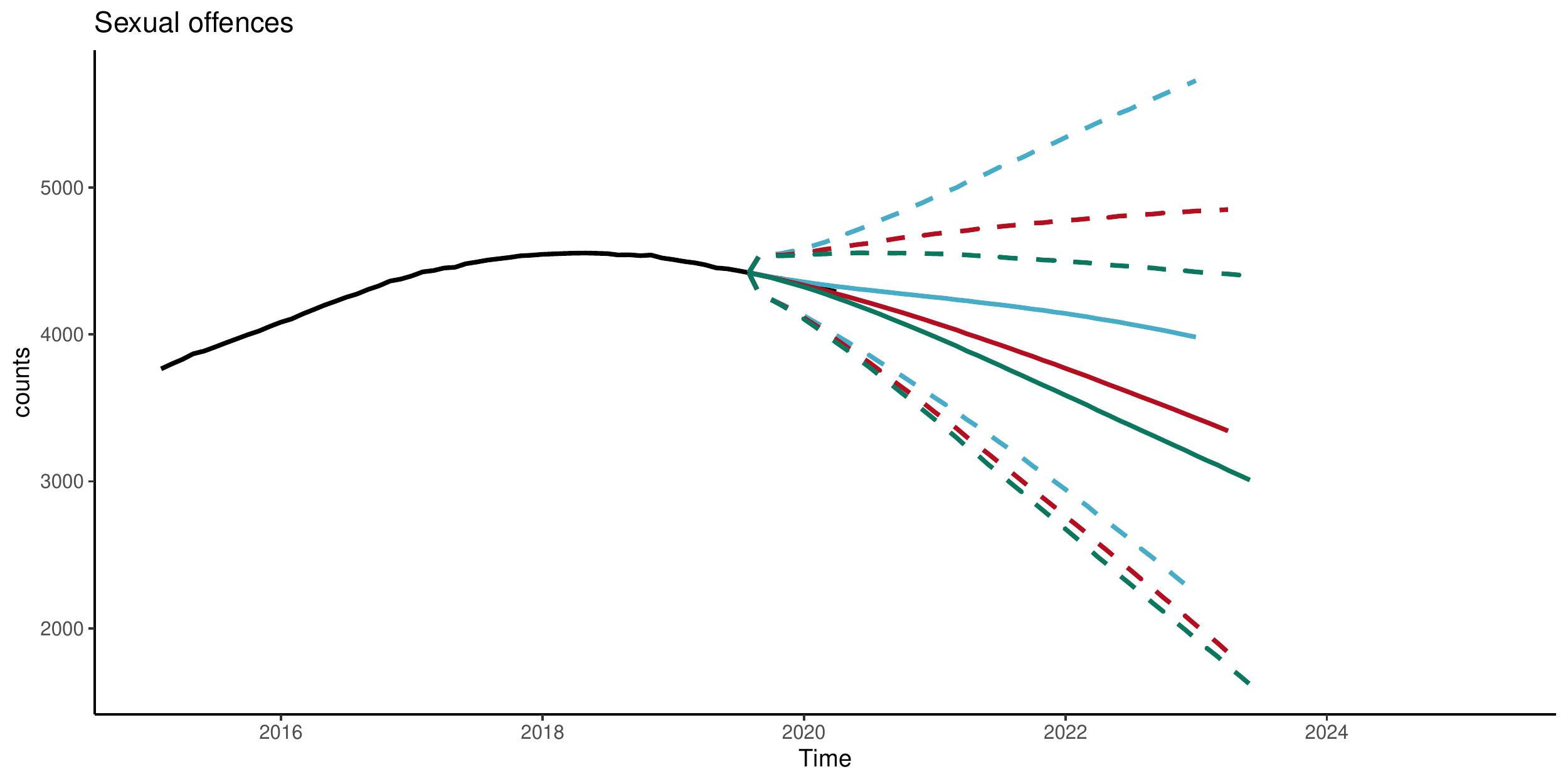}
\end{center}
\vspace{-1em}
\caption{Example \ref{ex:policymod}. Sentencing policy modification for Sexual offences group where $E[S^o]=29.71$. The red line is associated with unchanged parameters, the blue line is associated with $E[S^\nu]=48$ and the green line with $E[S^\nu]=24$. Solid line is the predictor. Dashed lines represent two standard deviation prediction intervals.}
\label{fig:Sex_scenarios1}
\end{figure}

\begin{figure}[h!]
\begin{center}
\includegraphics[width=0.5\textwidth,height=0.25\textheight]{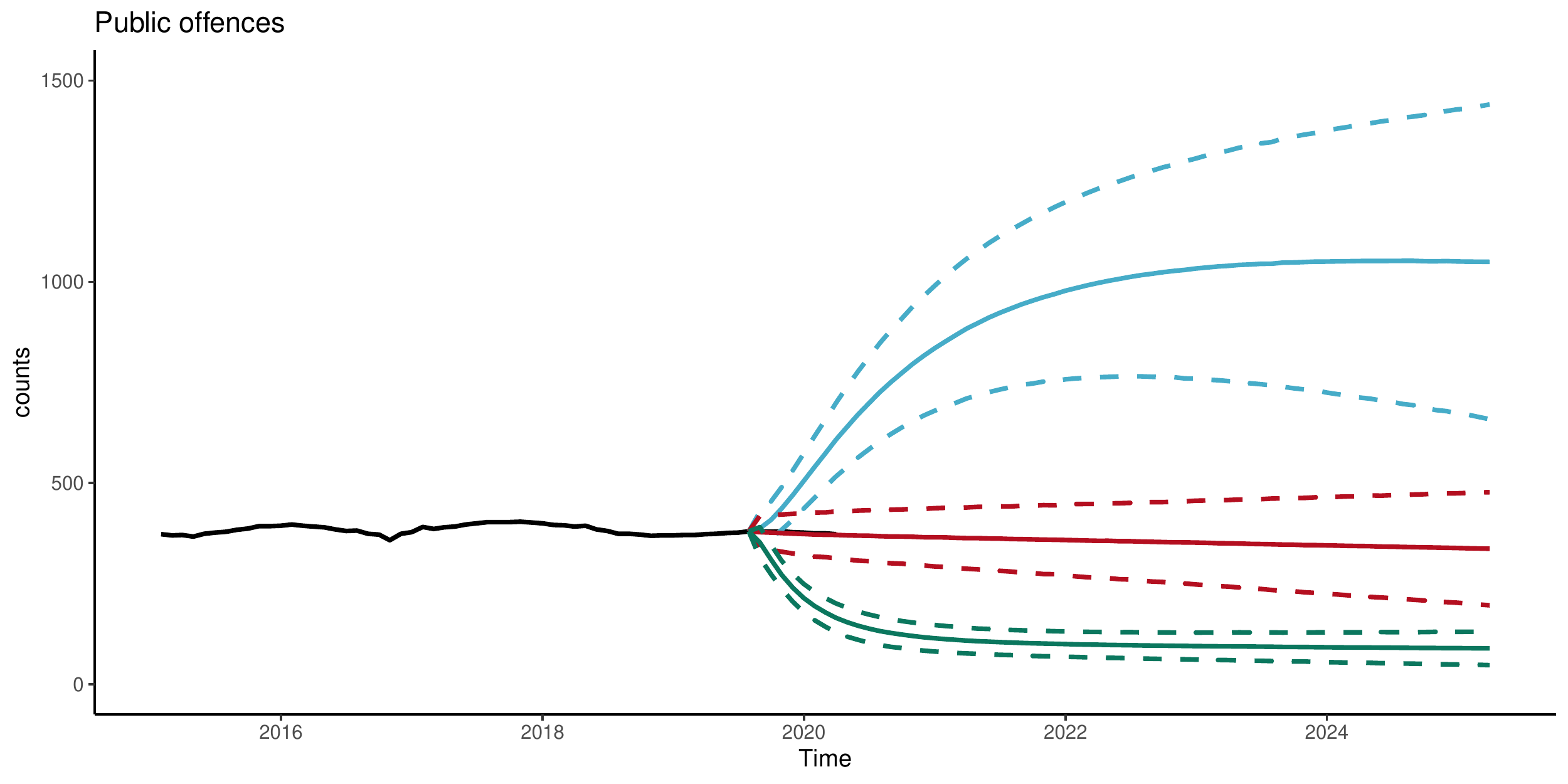}
\end{center}
\vspace{-1em}
\caption{Example \ref{ex:policymod}. Sentencing policy modification for Public order offences group where $E[S^o]=3.79$. The red line is associated with unchanged parameters, the blue line is associated with $E[S^\nu]=12$ and the green line with $E[S^\nu]=1$. Solid line is the predictor. Dashed lines represent two standard deviation prediction intervals.}
\label{fig:PublicOrder_scenarios1}
\end{figure}

\begin{example}\label{ex:policymod}
 A change point can be studied,  representing switching from one kind of service to another at time $\tau$ (that does not involve an observation), 
 the model  is an $M_t/G^o,G^{\nu}/\infty,$ \citep{Aras2017} (using the superscript $o$ for old and $\nu$ for new). 
The distribution of $Q(\tau+\delta)$ is Poisson with mean  $\hat{m}(\tau)  G^{o,c}_e + \check{m}(\tau+\delta)$.

In Figures \ref{fig:Theft_scenarios1}, \ref{fig:Sex_scenarios1} and \ref{fig:PublicOrder_scenarios1} we simulate the effect of increasing and decreasing the mean service time on the offence group population for theft offences, sexual offences and public order offences, respectively, and where the solid lines are the predictors and the dashed lines represent two standard deviation prediction intervals. We remark that to produce these results we reuse samples from the posterior distributions of $\beta_0, \beta_1$ and $\alpha$. The new values of $\theta$ are given by $\theta=E[S^{\nu}](\alpha-1)$. 

 \end{example}

\subsection{Discussion}\label{subsec:assumptions}

In the previous section we illustrated how to use the Bayesian model for short and longer-term predictions, and to provide insight into the implications of possible policy modifications. In Appendix \ref{app:steadystate}, drawing on other theoretical results, we briefly describe how modification of sentencing and custody rates can be seen as a form of intervention to enable congestion event recovery. Empirical demonstration of the value of such analysis is beyond the scope of the paper, as this would require data not available to us.

With regard to the predictions in Section \ref{subsec:sims}, we note that time-series forecasting methods such as ARIMA models \citep{Shumway2000} can predict the future prison population by describing the autocorrelation in the data. A direct comparison, using the data from Section \ref{subsec:Bayes}, is presented in Appendix \ref{app:comparison}, and in Section \ref{concludingremarks} we provide some comments on the different approaches to forecasting.

We now briefly discuss the impact of the assumptions of the mathematical model described in Section \ref{observedqueuesection}. It is assumed that the time served by individuals at the observation point $t=0$ is unknown, but if the elapsed service times  $\{y_{i}: i = 1, \dots, n \}$ of the observed population $\hat{Q}(t)$ are recorded, it can be more effective to use this information to make predictions, while carrying a greater cost \citep{Duffield1997}.  In contrast to Equation \eqref{ccdfremain}, the conditional remaining service time ccdf for elapsed service time $x$ is  
$ H^c_x(t)=  G^c(t+x)/G^c(x)$.

As the conditional remaining service times are no longer identically distributed, the complication of the distribution $\hat{Q}(t)$ increases and the resulting process is a Markov process \citep{Aras2017}, although the estimates for the mean number remaining at time $t$ and variance have simple forms:
$E[\hat{Q}(t)] = \sum_{i=1}^n H_{y_i}^c(t),  Var[\hat{Q}(t)] = \sum_{i=1}^n H_{y_i}^c(t) H_{y_i}(t)$.

  Further, as noted by \cite{Whitt1999c}, the importance of conditioning upon the time served at an observation point increases as the service-time distribution differs more from an exponential distribution. 
  If $G$ is highly variable, then the elapsed holding time can greatly help in future prediction  and a long elapsed holding time tends to imply a very long remaining holding time. Let $Y(\alpha, \theta) $ denote the Pareto service time distribution $G$  as defined in Example \ref{ex:pareto} and let $Y_x(\alpha, \theta) $ the associated ccdf for elapsed service time $x$. By the result for $H_x^c(t)$ above, \citet[Theorem 8]{Duffield1997} showed that $H_x^c(t) = G^c({t}(\frac{x}{\theta}+1)^{-1})$, which implied
$Y_x(\theta, \alpha) \stackrel{{d}}{=} \left(1+\frac{x}{\theta} \right)Y(\theta, \alpha)$.
Hence $E[Y_x(\theta, \alpha)] =  \left(1+\frac{x}{\theta}  \right) E[Y(\theta, \alpha)]$,
that is, the mean remaining service time $E[Y_t(\alpha, \theta)]$ is approximately proportional to the elapsed service time $t$.

\section{Concluding remarks}\label{concludingremarks}

This work was motivated by the problem of predicting the population of housed inmates within the prison system in England and Wales, where the size of the prison population is recorded on a regular basis, with attention both to short-term predictions to enable local resource planning, and longer-term projections that can provide insight to policy makers regarding the impact of potential policy variations. 

We studied the transient behaviour of the time-varying infinite server queue, $M_t/G/\infty$, fed by a non-homogeneous Poisson arrival process whose occupancy is observed at discrete points in time, but the time in service to that point is not known.
Drawing on this analysis, and using publicly available data, we built a model that could be used
as a decision support tool for custodial elements of the prison system. We illustrated the use of such a model for population prediction and for analysing the implications of changing external factors influencing the prison population such as  government guidelines and sentencing policies. The proposed model produces predictions together with uncertainty bands and aligns with the current guidelines on informed decision making in the UK government \citep{Aqua2015}.  

It is beyond our scope to compare the queueing theory approach to generating predictions with multiple forecasting methods, but we note recent work arguing that for time-series methods \cite{liu2021forecast} to support general scenario analysis requires extracting components known as `features' from the time-series data, that in turn can be used to generate alternative scenarios \citep{kegel2017whatif,tuominen2022forecasting}. In the queueing theory approach,  model parameters have a direct interpretation for the application domain, which straightforwardly enables the study of a variety of public policy initiatives by setting different input values to the model. We demonstrated this
in Example \ref{ex:policymod} for changes in the distribution of sentence lengths. As a further example, changes under consideration to the sentencing and release of serious and dangerous sexual and violent offenders (MoJ, \citeyear{PrisonProjections2019}), could be studied by changes in the arrival rate (e.g., the custody rate, conviction rate), studies we were unable to undertake, as they require data not available to us.
In contrast, time-series methods  are not so amenable to what-if style scenario analysis as they offer the possibility of forecasting future observations, but with limited interpretability of the fitted model \citep{petris2009}.

 The contributions of this paper are: (i) the novel synthesis of results from several authors about transient and stationary behaviour of the $M_t/G/\infty$ queue to enable construction of a model suited to short and longer-term predictions, and to supporting considerations of parameter uncertainty; and (ii) illustration of the approach to potential policy changes in the real-world domain of prison occupancy.

Reflecting the data available for model building, we focused on the situation where the system has non-empty initial state and where the elapsed time of each individual in the system is not known. The dynamics of the queue is a combination of those already in service at some time and those who subsequently arrive, and separation into initial content and new input allowed these sub-populations to be analysed jointly and separately. We drew on results for the transient and stationary distributions of these queueing systems from several authors to enable an analytic approach. 
Then, using a Bayesian approach with public historical data that allows the inclusion of expert knowledge,  we considered parameter uncertainty involved in the prediction of future arrivals and presented a model that maintains interpretability for the domain application. Incorporating other sources of uncertainty into our process \cite{Whitt2002}, including model and process uncertainty, and quantifying the contribution from each within our application is a topic of future research.

Further, we note that restoration of the departure process as approximately Poisson also allows the approach to be extended to a network of processes, referred to as a $(M_t/G/\infty)^N/M$ model,  in which queue length distribution models have time-dependent product form and would be appropriate for models of many service systems \citep{Massey1993}, including further development of a model specific to the prison domain.

 The approach is potentially applicable to other service systems, but the queueing model properties of interest will vary according to the application context, and the choice of what approach to take will also depend on the available data.  Data driven development of a system model is an increasingly popular approach as discussed in \cite{Mandelbaum2019}. However, the infrastructure to collect and manage data at multiple phases and timescales is not yet comprehensive, so parameter inference remains a challenging problem because of limited data availability about successive system states, and for model building methods that can take advantage of available but incomplete data are essential. 
 We acknowledge that some of our articulated modelling assumptions may not apply in  domains where fine-grained staffing implications are of interest, for example, hospitals and call centres.

The availability of future information (via predictive algorithms, machine learning, or local observation) and the resultant novel queueing analysis, has policy design implications, described as ``a broader shift from being reactive to proactive'' as future information becomes part of the policy maker's toolkit \citep{Spencer2014,Walton2021}. 

 \section*{Acknowledgements}
We would like to thank our colleagues from the Ministry of Justice for helpful discussions in the development of this work, and the journal reviewers for constructive comments that helped to improve the presentation. The work was conducted as part of the \emph{Managing Uncertainty in Government Modelling} project supported by The Alan Turing Institute. This work was supported by the Additional Funding Programme for Mathematical Sciences, delivered by EPSRC (EP/V521917/1) and the Heilbronn Institute for Mathematical Research.

\bibliographystyle{apalike} 
\bibliography{library} 
 
 \appendix
 
\section{Prison occupancy}\label{app:prisons}

\subsection*{Phases of the prison system}\label{subsec:phases}

Following an arrest by the police and being charged with a crime,  individuals are classified within 12 offence main groups (Table \ref{offencegroupstable}), which naturally form different service classes. The police decide if a person can be released on bail, but may have to follow certain rules until a hearing at a court, or if they are held on {remand} in prison until a hearing at a court. If an individual is found guilty by a court, a custodial sentence can be imposed. A sentence can be suspended, or served in the community and the vast majority of sentenced offenders do not spend their whole sentence in custody. The sentenced population is made up of Determinate sentenced and  Indeterminate sentenced. Each sentence is marked as one of two types: a standard determinate sentence (SDS) or an extended determinate sentence (EDS). Most offenders are given an SDS, which means that half way through their sentence they are released from prison and serve the remaining time in the community. An EDS means an offender serves 2/3 of their sentence in prison before being considered for release. Sentenced offenders who have been released from prison before the end of their sentence can be recalled to custody for breach of release conditions.

Figure \ref{model1sketch} displays a simplified version of prisoner journeys through the custodial system, with the total prison population divided into three main holding phases (displayed percentages are as at June 2019) (MoJ, \citeyear{PrisonProjections2019}): (i) on {remand} (11\%), (ii) sentenced prisoners (79\%) and (iii) on {recall} (9\%). Prisoners within the licence phase are in the community.

 Arrivals to the \emph{Remand} phase, are recorded in two streams, \emph{Untried prisoners} and \emph{Sentenced prisoners}. Individuals remain in remand for some time according to a length of stay distribution $G_{1}$. Upon completion of the Remand phase, the model assumes all Sentenced prisoners proceed to the Sentenced population and a proportion of the Untried prisoners, with the remaining released.  An arrival stream from the courts $ \lambda_2(t) \ge 0$ brings individuals directly into the Sentenced population.  
All departures from the Sentenced population are released into the community, and are all under \emph{Licence}, for some time, during which they can be  \emph{Recalled} into custody.  If an individual is recalled, they re-enter the custodial system and this figure assumes a prisoner can only be recalled once. In practice, this is not the case, although the data available does not provide information on repeatedly recalled prisoners.

\subsection*{Data}

Individual prison establishments record details of individuals on a prison IT system and this data is collected as a central database managed by MoJ and the HM Prison and Probation Service (HMPPS)  to produce the various analyses of prison population. Statistics are regularly released as well as projections of the prison population \citep{OffenderManage2019, MoJStory2016}. 

Between 1993 and 2012 the prison population in England and Wales increased by 41,800 prisoners to over 86,000, studies of this growth can point to patterns in the data that align with earlier identifiable policy changes. Such examples illustrate how modelling approaches enabling an understanding of the impact of possible policy changes can be valuable in this domain.  

Since 1993 the courts sentenced more offenders each year. Almost all of the increase (98\%) took place within two segments of the population - those sentenced to immediate custody (85\% of the increase) and those recalled to prison for breaking the conditions of their release (13\% of the increase).  Three offence groups, violence against the person, drug offences and sexual offences have had a particular impact on the prison population: the number of these offenders increased and the mean length of stay increased. 

The proportion of offenders whose sentences were 4 years or more, grew to 54\% of the sentenced prison population in 2012, compared to 45\% in 1993.
The mean sentence length over all offences in 2018 was 17.3 months. For more serious offences, the mean prison sentence was 58.3 months, more than 24 months longer than in 2006. More than two and a half times as many people were sentenced to 10 years or more in 2018 than in 2006.

From 1999 to 2011, the mean time served increased from 8.1 to 9.5 months for those released from determinate sentences. This was likely due to an increase in the mean custodial determinate sentence length handed down by the courts, and a decline in the parole release rate which meant that offenders had served longer by the time they were released. The second largest increase was within the recall population, caused by changes to the law making it easier to recall prisoners, and changes introduced in the Criminal Justice Act 2003 which lengthened the licence period for most offenders.  
   
  \section{Steady state recovery from congestion}\label{app:steadystate}

To illustrate the applicability of further theory to support policy analysis,
we present \cite{Duffield1997}'s analysis of what is termed \emph{congestion} events, that is, where an unusually high number of individuals has been observed, and allowing one to investigate the effects of possible interventions (such as change of arrival rate, or pausing arrivals) on recovery from congestion. These results rely on an assumption that the system is in steady state, i.e., $\lambda(t) = \lambda$, which could be motivated by a slowly changing arrival rate, or for scenario analysis. These results follow directly from the results for the model described in Section \ref{subsec:conditional} and can be employed to characterise the effect of possible interventions that could be introduced to change the rate of recovery. 

For all $\tau$,  $\nu = \nu_\tau = \lambda E[S]$, $\check{m}(\tau+\delta) =  \lambda E[S] G_e(\delta) $ and  $p_{\tau}(\delta) = G_\tau^c(\delta)=  G_e^c(\delta)$ and the conditional mean is 
 \begin{align} \label{constant_mean}
 E \left [Q(\tau+\delta)  \mid Q(\tau) = n\right ]  &= \lambda E[S] G_e(\delta) +n G_e^c(\delta).
 \end{align}
As \cite{Goldberg2008} noted, since $ \nu_\tau $ and $G_\tau^c(\delta)$ can be computed, it can be directly determined when the steady state approximation is reasonable.
 
 \cite{Duffield1997} proposed the \emph{steady state rare event}, as an approximation for what is seen at a random (steady state) time and also as an approximation for the associated hitting time event. 
The likelihood of a high congestion event, that is, having a large number $N$ of individuals present, $P\left[N \ge n \right] = \sum_{k = n}^\infty P[Po(\nu) = k]$ with  $\nu = \lambda E[S]$ at $t=0$. 

Define a recovery level $k$, then a system has recovered from a rare congestion event at time $0$ when $Q(t)$ first reaches $k$ with $\nu <k<n$.  
 To define the recovery time for the mean, let
\begin{align} \label{meanrecoverydefn}
\beta \equiv \beta_{n,k} = \inf \left \{t \ge 0 : E \left [Q(t)  \mid Q(0) = n\right ]\le k \right\}. 
\end{align}
The following result describes how the system recovers from congestion and the way it depends on cdf $G$.

\begin{theorem}\label{th:recovery} \citep[Theorem 2]{Duffield1997} If there is no $t$ such that $G^c(t^-) >0 = G^c(t)$, then $G_e^c(t)$ is continuous and strictly increasing. Then the recovery time for the mean, $\beta$ in Equation \eqref{meanrecoverydefn} is the unique root of the recovery equation, 
\begin{align} \label{recoveryequation}
 G^c_e(t) = \frac{k-\nu}{n-\nu}.
 \end{align}
 More generally, $ \beta = \sup  \left \{t: G^c_e(t) > \frac{k-\nu}{n-\nu} \right \}$.
\end{theorem}

This result is obtained by rearranging Equation \eqref{constant_mean}. It can be seen that the impact of $G$ on $\nu$ depends on whether recovery level $k$ is close to $n$ or close to $\nu$. When the recovery level $k$ is close to $\nu$, the cdf matters greatly. If the recovery level $k$, is closer to $n$ than $\nu$, then the recovery time $\nu$ depends more on the initial portion of the cdf $G_e$ than upon its tail. Consistent with this observation, the impact of a long-tail cdf on $\nu$
differs little from the impact of a short-tail cdf with the same mean if $k$ is suitably close to $n$, but it differs dramatically if $k$ is suitably close to $\nu$. However, `suitably close' depends on the decay rate of the ccdf $G_e^c$.  
 
If the system does experience a congestion event, the recovery can be aided by reducing the arrival rate, or by pausing arrivals until the congestion has cleared. If the arrival rate is reduced to a new constant rate, is equivalent to reducing the mean, as $\nu = \lambda E[S]$, produces a new recovery equation of the form in Equation \eqref{recoveryequation}. For $k<m<n$, clearly $(k-\nu)/(n-\nu)$ is increasing in $\nu$, so that the recovery time for the mean is reduced by decreasing $\nu$. If the arrivals are terminated until recovery is achieved, $\nu=0$ in the recovery equation. Then further recovery can be achieved by turning the arrivals on and replacing $n$ with $k$, and $j$ for $k$ in the recovery equation, such that $m<j<k$.

\begin{figure}[h!]
   \centering
  \includegraphics[width=0.5\textwidth]{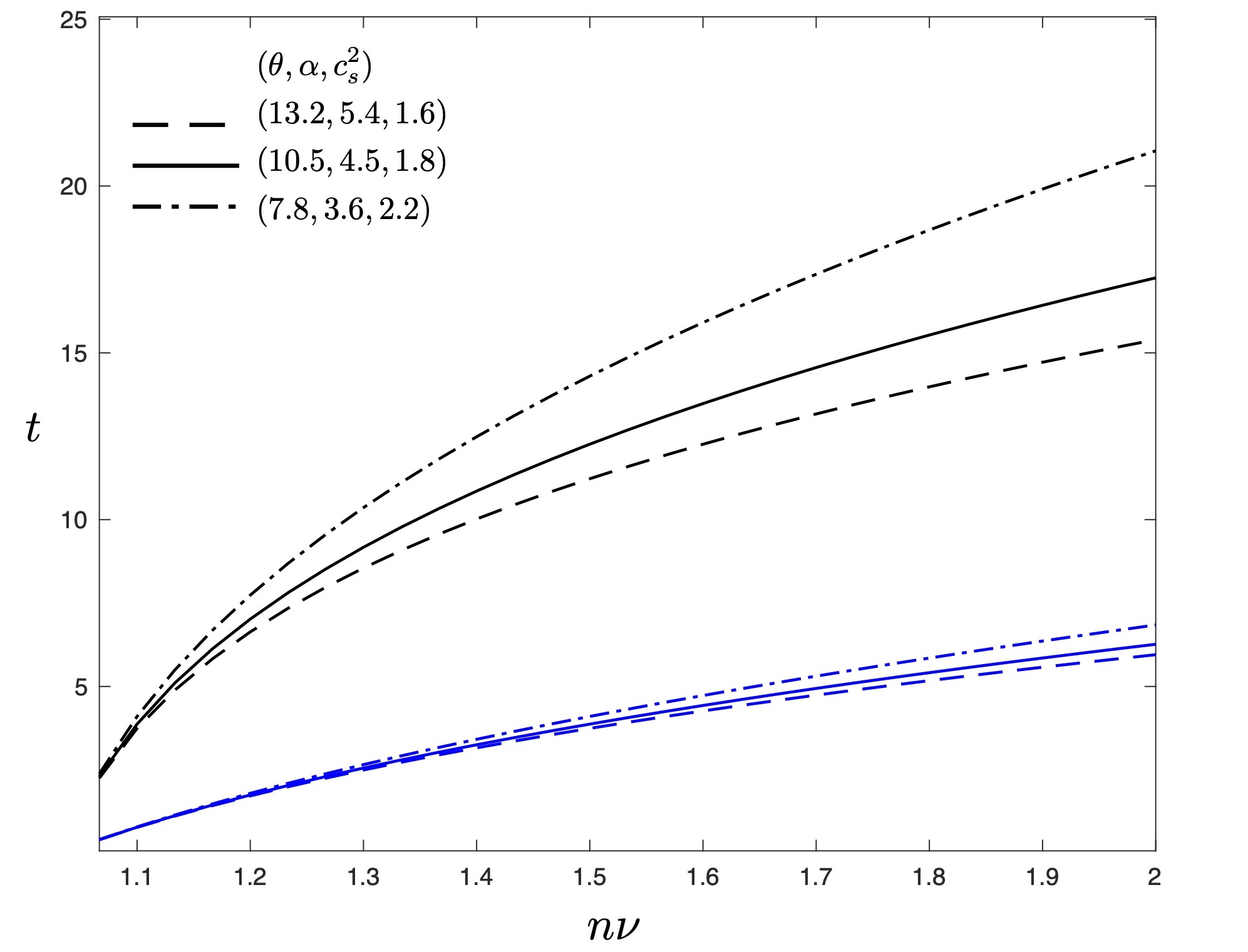}
\caption{Example \ref{ex:pareto_recovery}. Congestion level relative to the mean $n \nu$ vs. recovery time $t$. For fixed $E[S] = 3$, $\nu = 30$ and varying $(\theta, \alpha, c_s^2)$ (black lines). An intervention of  $\lambda_2 = 0.8 \lambda_1$ is illustrated by the corresponding blue lines.}
\label{fig:example_recovery}
\end{figure}

\begin{example}\label{ex:pareto_recovery} For $G \sim Pa(\theta,\alpha$),   the recovery equation in Equation \eqref{recoveryequation} we have $t =  \left (\frac{\theta^{\alpha-1}(n-\nu)}{k-\nu} \right)^{\frac{1}{\alpha-1}} -  \theta$. In Figure \ref{fig:example_recovery}, for fixed $E[S] = 3$  and recovery level $k =  \lceil \nu+1 \rceil$, the black lines illustrate the recovery time for varying levels of congestion for different scenarios $(\theta,\alpha, c_s^2)$, where the x-axis shows the congestion level relative to the steady state mean $\nu$. To aid the recovery for the three scenarios (black lines), we reduce the arrival rate by 20 percent, that is, $\lambda_2 = 0.8 \lambda_1$ and illustrate the effect on the corresponding dashed lines (blue lines).
\end{example}

\section{Comparison to time series analysis}\label{app:comparison}
\begin{figure}[h!]
\begin{center}
\includegraphics[width=0.65\textwidth,height=0.23\textheight]{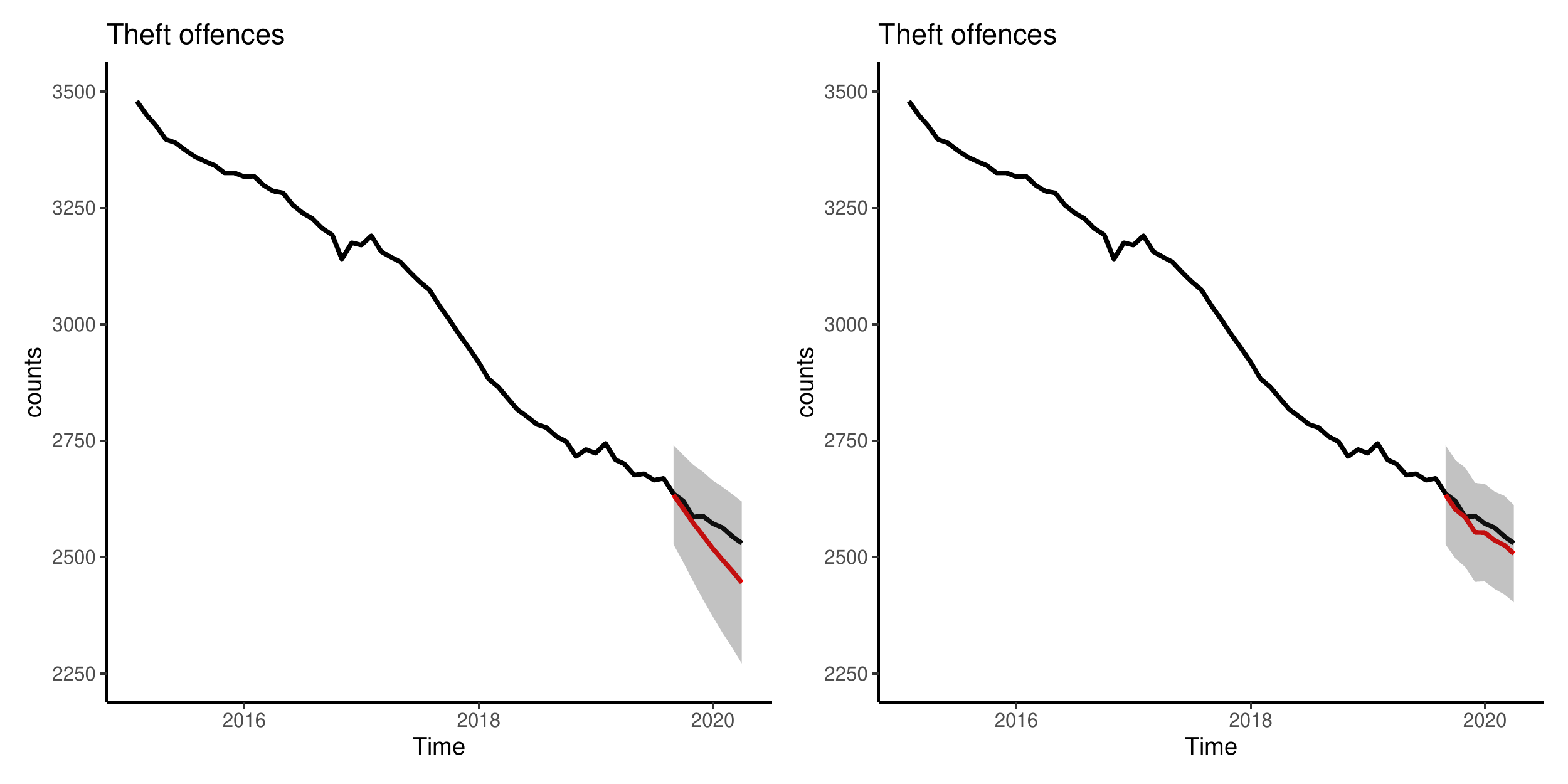}
\end{center}
\vspace{-1em}
\caption{Prediction of the number of prisoners in the Theft offences group from August 2019 to March 2020. \textit{Left panel:} long-term projections (red line) together with two standard deviation prediction interval (grey shaded region). \textit{Right panel:} short-term projections (red line) together with two standard deviation prediction interval (grey shaded region). The observed values are in black.}
\label{fig:Prediction_Theft}
\end{figure}

\begin{figure}[h!]
\begin{center}
\includegraphics[width=0.65\textwidth,height=0.23\textheight]{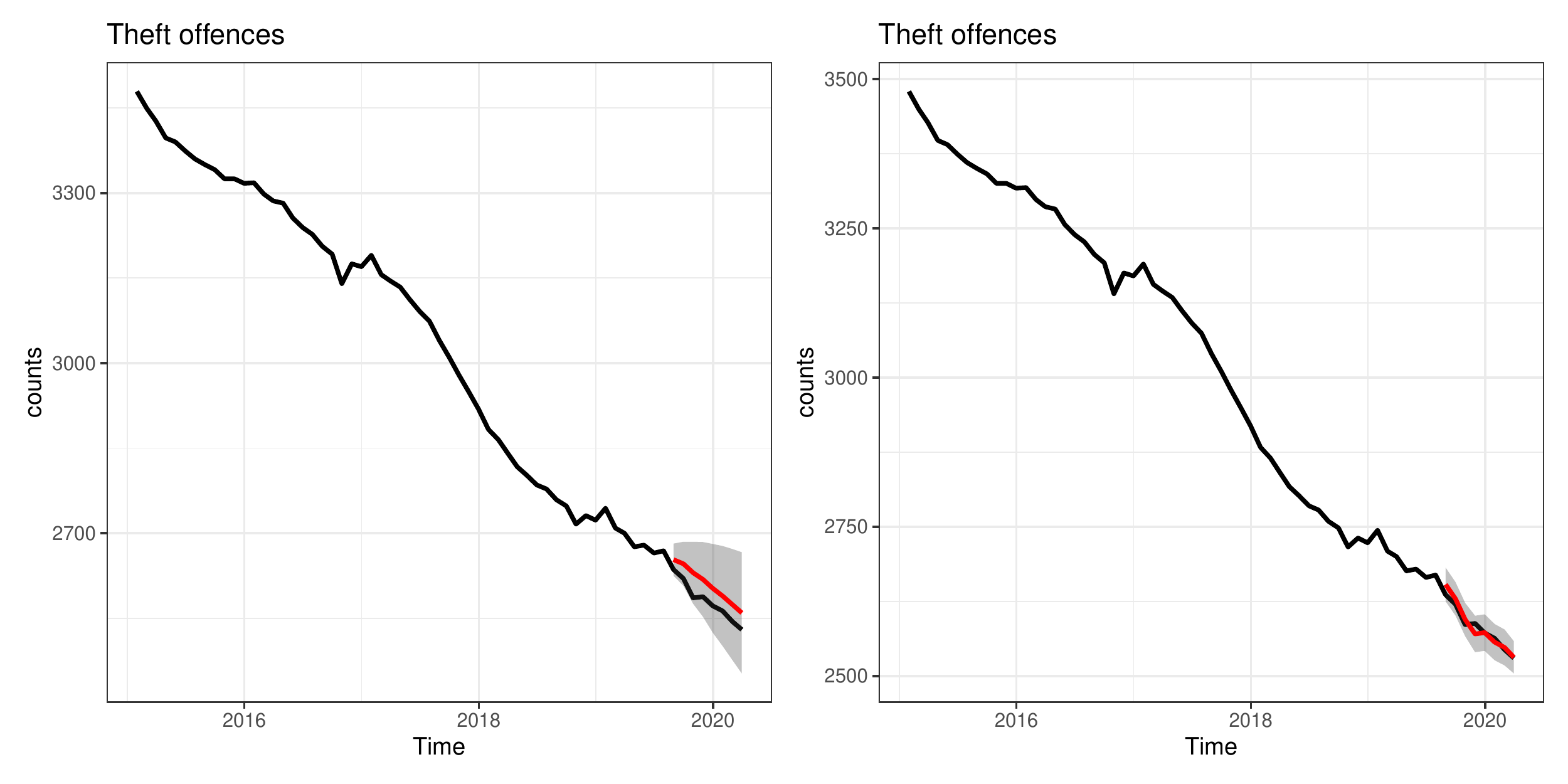}
\end{center}
\vspace{-1em}
\caption{ARIMA model prediction of the number of prisoners in the Theft offences group from August 2019 to March 2020. \textit{Left panel:} long-term projections (red line) together with two standard deviation prediction interval (grey shaded region). \textit{Right panel:} short-term projections (red line) together with two standard deviation prediction interval (grey shaded region). The observed values are in black.}
\label{fig:ARIMA_Prediction_Theft}
\end{figure}

\begin{figure}[h!]
\begin{center}
\includegraphics[width=0.65\textwidth,height=0.23\textheight]{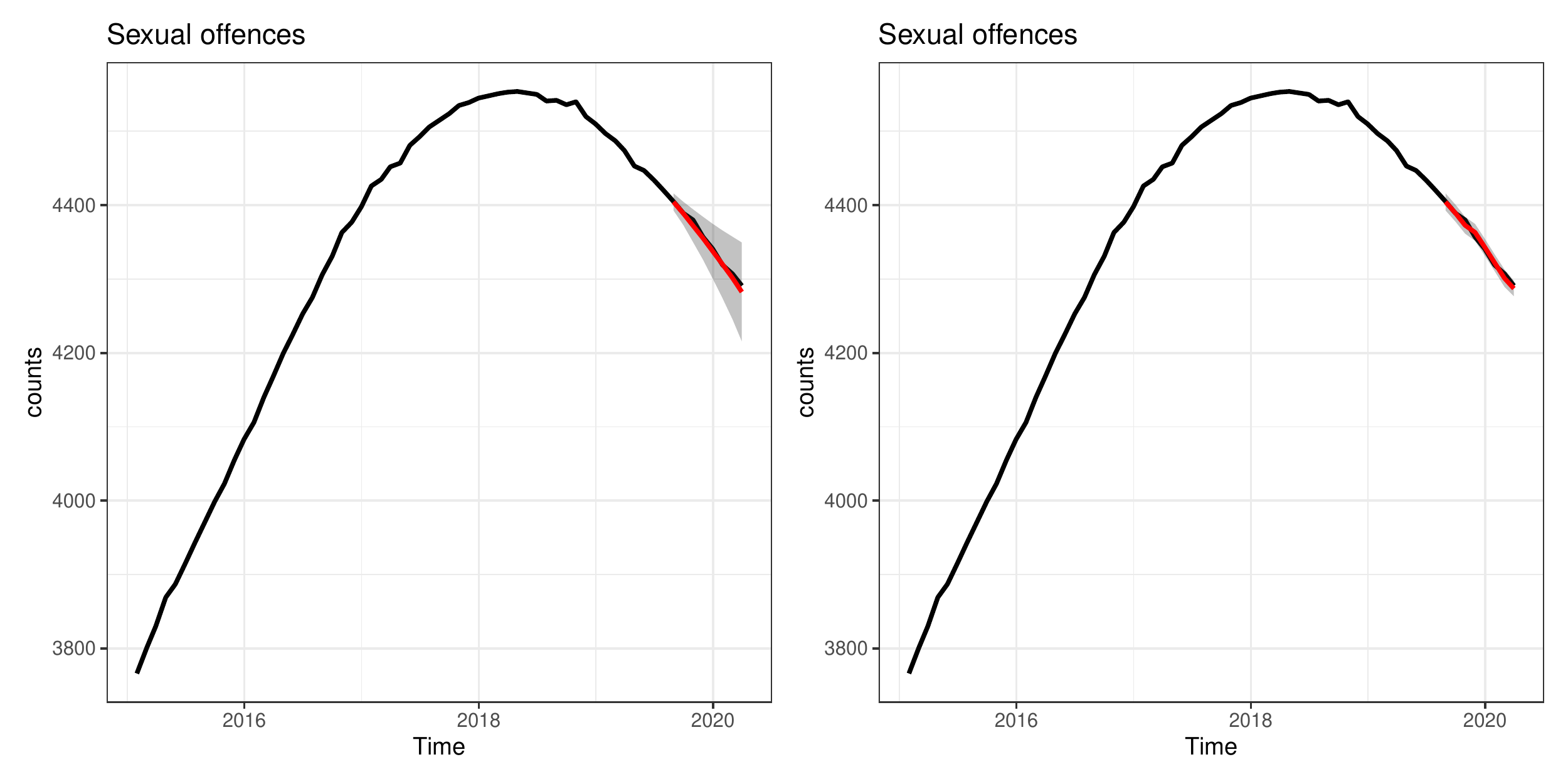}
\includegraphics[width=0.65\textwidth,height=0.23\textheight]{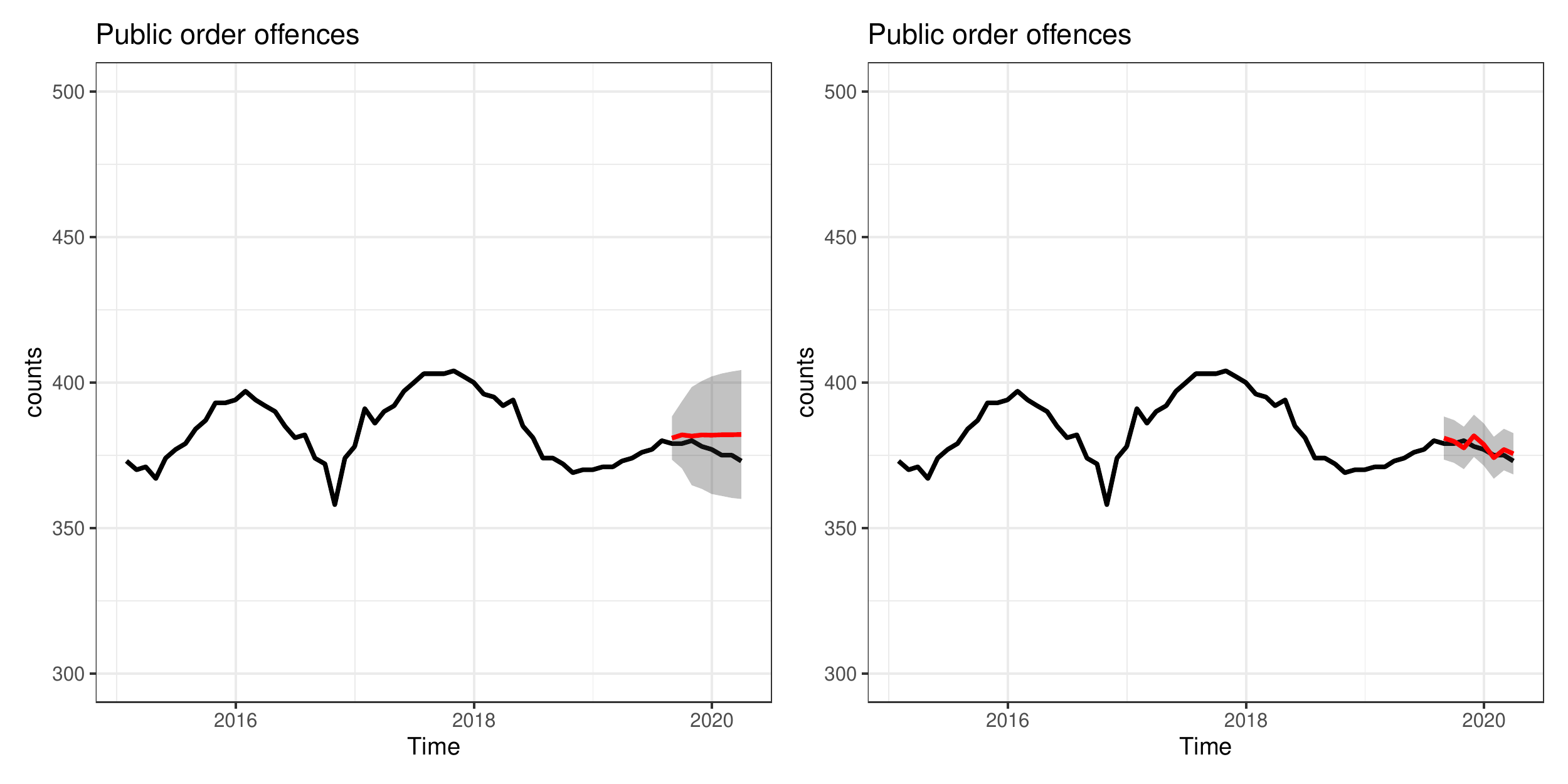}
\end{center}
\vspace{-1em}
\caption{ARIMA model prediction of the number of prisoners in the Sexual offences (\textit{top row}) and Public order offences (\textit{bottom row}) groups from August 2019 to March 2020. \textit{Left panel:} long-term projections (red line) together with two standard deviation prediction interval (grey shaded region). \textit{Right panel:} short-term projections (red line) together with two standard deviation prediction interval (grey shaded region). The observed values are in black.}
\label{fig:ARIMA_Prediction_PublicOrder}
\end{figure}

We apply traditional time-series methods, ARIMA models \citep{Shumway2000} to produce prison population predictions for the offence groups considered in Section \ref{subsec:sims}.  The order and the values of parameters are obtained using \texttt{R} package \texttt{forecast} \citep{Hyndman2008}. The forecasts of the number of prisoners in the Theft offence group are produced by ARIMA(2,1,0) with drift given by:
\begin{align*}
    y_t&=-9.76 +0.94y_{t-1}+0.462y_{t-2} - 0.41y_{t-3} +\epsilon_t.
\end{align*}
Similar to the analysis performed in Section \ref{subsec:sims}, one-step ahead predictions are produced by iteratively adding one data point at a time and refitting the model and we note that the order of the model was updating with the changes in the training data set. The RMSE for long-term predictions is 30.21, whereas for short-term predictions is 10.16. The results are presented in Figure \ref{fig:ARIMA_Prediction_Theft}, and we observe that the length of the prediction intervals is lower compared to the predictions produced in Figure \ref{fig:Prediction_Theft}. This is due to the fact that the parameters of ARIMA model are set to the fixed quantities, whereas we assign probability distributions for the parameters in our proposed model. Similar results are presented  for the Sexual offences and Public order offences groups in Figure \ref{fig:ARIMA_Prediction_PublicOrder}. 
For example we present the $k$-step ahead forecasts for the Sexual offences group which are produced by ARIMA(2, 2, 2):
\begin{align*}
y_t &=2.55y_{t-1}-1.95y_{t-2}+0.23y_{t-3}+0.16y_{t-4}+\epsilon_t-1.62\epsilon_{t-1}+0.68\epsilon_{t-2},
\end{align*}
where the RMSE for long-term predictions is 4.71, whereas for short-term predictions is 4.42, where the above comment holds observing lower prediction intervals. The $k$-step ahead forecasts for the Public order offence group was produced by ARIMA(2,0,2):
\begin{align*}
    y_t&=93.22 + 0.28y_{t-1} + 0.47y_{t-2}+\epsilon_t+0.91\epsilon_{t-1} + 0.85\epsilon_{t-2}, 
\end{align*}
where the RMSE for long-term predictions is 5.44, whereas for short-term predictions is 2.16.

\newpage

 \begin{table}[h!]
\caption{Offence groups}    
\vspace{-2mm}
\begin{center}
\begin{small}
\begin{tabular}{clcl}
\toprule
1.&Violence against the person&7. & Possession of weapons\\
2. &Sexual offences&8. & Public order offences\\
3.& Robbery&9. & Miscellaneous crimes against society\\
4.& Theft Offences&10. & Fraud Offences\\
5. & Criminal damage and arson&11. & Summary Non-Motoring\\
6. & Drug offences&12. &Summary motoring\\
\hline
\end{tabular}
\end{small}
\end{center}
\label{offencegroupstable}
\end{table}%

\end{document}